\documentclass[aps,prl,twocolumn,superscriptaddress,showpacs,floatfix]{revtex4-1}

\usepackage{amsmath,amssymb,amsfonts,float,graphics,epsfig,epstopdf,color,verbatim,tabularx,bm,multirow,appendix}
\usepackage[utf8]{inputenc}
\usepackage[T1]{fontenc}
\usepackage{xcolor}

\usepackage{dsfont}
\usepackage{textcomp}
\usepackage{yfonts}
\usepackage{bm}
\usepackage{subfigure}
\usepackage{mathrsfs}
\usepackage{graphicx}
\usepackage{verbatim}
\usepackage{hyperref}
\usepackage{multirow}
\usepackage{braket}
\usepackage[normalem]{ulem}
\usepackage{tikz}
	\usetikzlibrary{calc}
	\usetikzlibrary{shapes.multipart}

\newcommand{\comments}[1]{}

\newcommand{\stkout}[1]{\ifmmode\text{\sout{\ensuremath{#1}}}\else\sout{#1}\fi}

\makeatletter
\def\l@subsubsection#1#2{}
\makeatother

\begin{document}

\title{Scaling of entanglement entropy at deconfined quantum criticality}

\author{Jiarui Zhao}
\affiliation{Department of Physics and HKU-UCAS Joint Institute of Theoretical and Computational Physics, The University of Hong Kong, Pokfulam Road, Hong Kong SAR, China}

\author{Yan-Cheng Wang}
\affiliation{School of Materials Science and Physics, China University of Mining and Technology, Xuzhou 221116, China}

\author{Zheng Yan}
\affiliation{Department of Physics and HKU-UCAS Joint Institute of Theoretical and Computational Physics, The University of Hong Kong, Pokfulam Road, Hong Kong SAR, China}
\affiliation{State Key Laboratory of Surface Physics and Department of Physics, Fudan University, Shanghai 200438, China}

\author{Meng Cheng}
\email{m.cheng@yale.edu}
\affiliation{Department of Physics, Yale University, New Haven, CT 06520-8120, U.S.A}

\author{Zi Yang Meng}
\email{zymeng@hku.hk}
\affiliation{Department of Physics and HKU-UCAS Joint Institute of Theoretical and Computational Physics, The University of Hong Kong, Pokfulam Road, Hong Kong SAR, China}

\begin{abstract}
We develop a nonequilibrium increment method to compute the R\'enyi entanglement entropy and investigate its scaling behavior at the deconfined critical (DQC) point via large-scale quantum Monte Carlo simulations. To benchmark the method, we first show that at an conformally-invariant critical point of O(3) transition, the entanglement entropy exhibits universal scaling behavior of area law with logarithmic corner corrections and the obtained correction exponent represents the current central charge of the critical theory. Then we move on to the deconfined quantum critical point, where although we still observe similar scaling behavior but with a very different exponent. Namely, the corner correction exponent is found to be negative. Such a negative exponent is in sharp contrast with positivity condition of the R\'enyi entanglement entropy, which holds for unitary conformal field theories(CFT). Our results unambiguously reveal fundamental differences between DQC and quantum critical points(QCPs) described by unitary CFTs.
\end{abstract}
\date{\today}
\maketitle

{\it{Introduction.-}}
Quantum many-body entanglement has become a fundamental organizing principle for the study of quantum matter. Scaling behavior of entanglement entropy (EE) provides deep insights into the structure of quantum many-body states and gives universal invariants that can be used to characterize distinct phases and phase transitions. For these reasons
 EE has been of interests to many, ranging from the field theoretical to numerical and experimental communities of quantum many-body systems~\cite{Calabrese_2004,Fradkin2006,Casini2006,Kitaev2006,Levin2006,Hastings2010,Metlitski2011,Isakov2011,Jiang2012,Casini2012,Swingle2012,Humeniuk2012,Inglis2013,InglisNJP2013,KallinPRL2013,Luitz2014,KallinJS2014,Helmes2014,Laflorencie2016,PhysRevLett.96.010404}. For $(2+1)$d quantum critical points, the EE obeys the ``area law'', i.e. linearly proportional to the perimeter of the entangling region. However, the subleading term turns out to be more interesting~\cite{Calabrese_2004,Fradkin2006,Casini2006,Kitaev2006,Levin2006,PhysRevLett.96.010404,PhysRevLett.99.147202,PhysRevB.77.140402,PhysRevB.86.214203,ref12}: it is either a universal constant when the entangling region has a smooth boundary, or a logarithmic term with a universal coefficient when the boundary contains sharp corners. The corner correction has been shown to be deeply related to intrinsic data of the underlying conformal field theory. For example, the universal coefficient for the von Neumann EE in the smooth limit is essentially given by the stress tensor central charge of the CFT. On the other hand, the scaling form of EE has also been investigated in numerical simulations of microsopic lattice models. In particular, the corner corrections were also identified in quantum Monte Carlo (QMC) simulations of QCPs for conventional symmetry-breaking transitions~\cite{InglisNJP2013,KallinPRL2013,KallinJS2014,Helmes2014,Laflorencie2016,JRZhao2020}, the results of which are largly consistent with field-theoretical predictions.

For QCPs beyond the paradigm of Landau-Ginzburg-Wilson, the scaling forms of EE are not well understoond. Among these, the quantum entanglement of the deconfined quantum criticality (DQC)~\cite{Sandvik2007,Senthil2004,NSMa2018,Liu_2019,Jiang2008,Banerjee:2013dda} -- a continuous quantum phase transition between two seemingly unrelated symmetry-breaking states -- has not been explored much. Theoretically, since the proposed low-energy theory of DQC is a strongly coupled gauge theory~\cite{Senthil2004,Swingle2012,CWang2017,DCLu2021}, no controlled analytical treatment is available. While conventional O$(n)$ CFTs are also interacting, they turn out to be ``close'' to the free Gaussian theory, e.g. the corner correction is well estimated by the value of the free theory.  For DQC, basic question such as whether the generic scaling form from CFT still holds is not known. In this work, we will address these questions using large-scale, unbiased QMC simulations.

\begin{figure}[htp]
	\centering
	\includegraphics[width=\columnwidth]{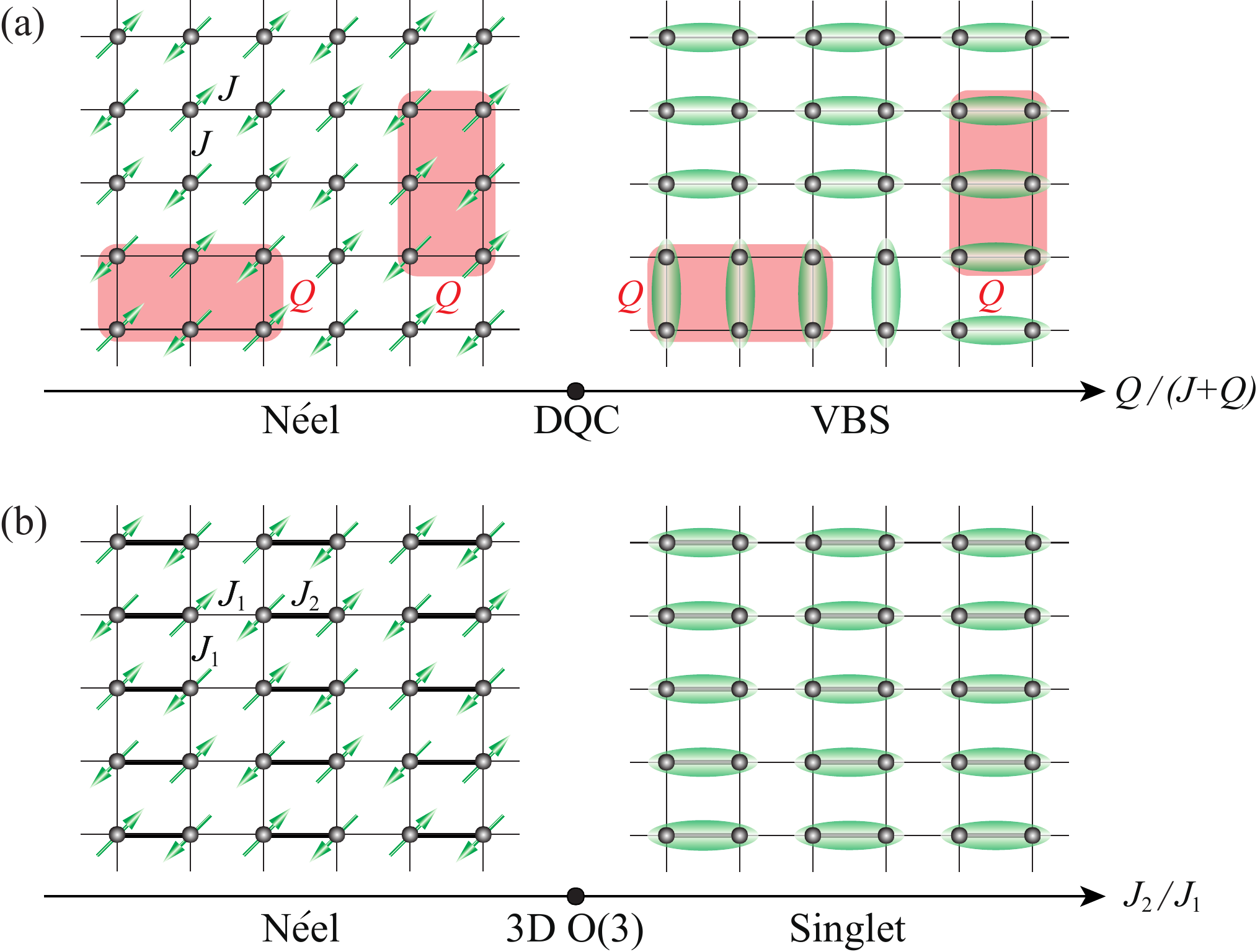}
	\caption{The two lattice models: (a) the $J$-$Q_3$ model which hosts DQC~\cite{Lou2009}, (b) the square lattice $J_1$-$J_2$ antiferromagnetic columnar dimer model which exhibits $(2+1)d$ O(3) QCP as $J_2/J_1$ is tuned~\cite{NSMa2018anomalous}.}
	\label{fig:fig1}
\end{figure}

Because EE is a non-local quantity, its numerical computation, even just the second R\'enyi EE, is a challenging task in QMC simulations of interacting lattice models.  Although several numerical algorithms have been developed for this purpose~\cite{Hastings2010, Humeniuk2012,Inglis2013,Luitz2014}, they are still numerically very costly due to the necessity of enlarging the configurational space to replicas and exchanging them during the sampling processes. Further improving the efficiency and stability of the numerical estimator, especially for large system sizes and lattice models with multi-spin interactions or fermions~\cite{Grover2013,Assaad2015} is still highly desirable.

These are the difficulties we set out to overcome in this work. Building on the recent nonequilibrium measurement of the R\'enyi entropy~\cite{DEmidio2020} which has shown its unprecedented efficiency on the measurememt of R\'enyi entanglement entropy of two dimensional Heisenberg model than the previous attempts ~\cite{PhysRevLett.104.157201,PhysRevB.83.224410,PhysRevB.84.165134,PhysRevB.86.235116,PhysRevA.102.062413}, we develop a new nonequilibrium increment method, that can make the best usage of the divide-and-conquer procedure of the nonequilibrium process and the modern massive parallel computing technique to improve the speed of the simulation and the data quality of the entanglement measurement~\cite{JRZhao2021method}.

To test the performance of our method, we first show that at the $(2+1)d$ O(3) transition in a square lattice $J_1-J_2$ columnar dimer model, the EE indeed exhibits universal scaling behavior of area law plus logarithmic corner corrections, and the obtained correction exponent is closer to the prediction of Gaussion theory ~\cite{CASINI2007183} consistent with previous numerical results~\cite{KallinJS2014,Helmes2014,InglisNJP2013}. Then we move on to the DQC and find that although the EE still obeys a similar scaling behavior, the universal coefficient of the corner correction term is negative.  Such a result is in sharp contradiction with the positivity conditions for the R\'enyi EE which hold for unitary conformal field theories~\cite{Casini2012,Bueno_2015, YCWang2021DQCdisorder} -- and pointing towards alternative scenarios of DQC such as non-unitary CFT with complex fixed points annihilation~\cite{Nahum2020,RCMa2020,YCWang2021DQCdisorder},  multi-criticality~\cite{BWZhao2020PRL,DCLu2021} or precursors to a weakly first-order transition~\cite{Jiang2008,Kuklov2008,KChen2013,DEmidio2021}.

{\it{Model.-}}
Our main goal is to investigate the 2nd R\'enyi EE at the DQCP of the $J-Q_3$ model~\cite{Sandvik2007,Senthil2004,NSMa2018}, as illustrated in Fig.~\ref{fig:fig1} (a). The Hamiltonian reads
\begin{equation}
	H_{J-Q_{3}}=-J\sum_{\langle ij \rangle}P_{i,j}-Q\sum_{\langle ijklmn \rangle}P_{ij}P_{kl}P_{mn}
\end{equation}
where $P_{ij}=\frac{1}{4}-\mathbf{S}_{i}\cdot\mathbf{S}_{j}$ is the two-spin singlet projector. The quantum critical point separating the antiferromagnetic N\'eel and valence bond solid (VBS) states is at $[Q/(J+Q)]_c=0.59864(4)$~\cite{Lou2009,Shao2016,YCWang2021DQCdisorder}.

We also investigate the square lattice columnar dimer model, shown in Fig.~\ref{fig:fig1} (b). The Hamiltonian is given by
\begin{equation}
	H_{J_1-J_2}=J_1\sum_{\langle ij \rangle} \mathbf{S}_{i}\cdot\mathbf{S}_{j}+J_2\sum_{\langle ij \rangle^{'}} \mathbf{S}_{i}\cdot\mathbf{S}_{j},
	\label{eq:J1J2H}
\end{equation}
where $\langle ij \rangle$ denotes the thin $J_1$ bond and $\langle ij \rangle^{'}$ denotes the thick $J_2$ bond and the QCP $(J_2/J_1)_c=1.90951(1)$~\cite{NSMa2018anomalous} is known to fall within the $(2+1)d$ O(3) universality class. As explained in the Supplemental Material (SM)~\cite{suppl}, because of the translation symmetry breaking due to the strong $J_2$ and weak $J_1$ bonds in Eq.~\eqref{eq:J1J2H}, the entangling region $A$ must be chosen so that its boundary avoids strong dimer bonds to correctly extract the scaling behavior of EE from finite-size data. 

At a conformally-invariant QCP, the $n$-th R\'enyi EE of an entangling region $A$ of linear size $l$ is expected to take the following form:
\begin{equation}
S^{(n)}_A(l)=a_nl-s_n\ln l+b_n.
\label{eq:REE}
\end{equation}
Here $s_n$ is a universal constant of the underlying CFT, which only depends on open angles of the sharp corners of $A$: $s_n=\sum_{j}s_n(\alpha_j)$, where $\alpha_j$ is the open angle~\cite{Calabrese_2004,Fradkin2006,Laflorencie2016}. Analytical results about the universal function $s_n(\alpha)$ are only available in the extreme cases $\alpha\rightarrow 0$ and $\alpha\rightarrow \pi$. In addition, one can prove that generally in a unitary CFT $s_n(\alpha)$ must be non-negative and is a concave function of $\alpha$~\cite{Casini2006,Casini2012}.  Numerically, the corner correction has been systematically investigated in $(2+1)$d O$(n)$ models~\cite{InglisNJP2013,KallinJS2014,KallinPRL2013,Helmes2014,Laflorencie2016,JRZhao2020}.  Our goal is to extract $s$ for 2nd R\'enyi entropy at DQC for $\alpha=\pi/2$ since we will only consider rectangle regions.

\begin{figure}[htp]
\centering
\includegraphics[width=\columnwidth]{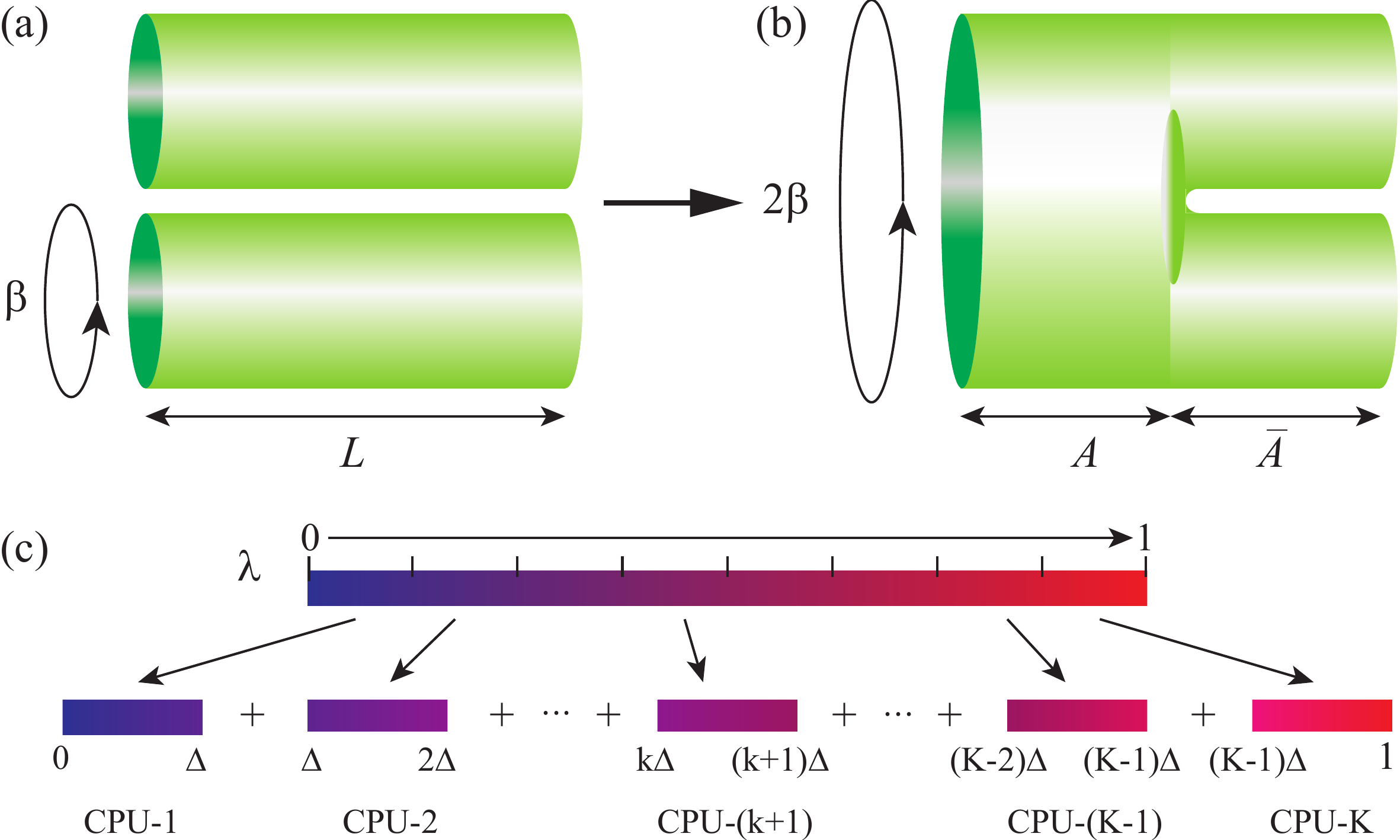}
\caption{The schematic diagram of the QMC configurations and the nonequilibrium increment method. (a) The QMC configuration of $\mathcal{Z}_{\varnothing}^{(2)}$. The configuration is two independent replicas with periodical boundary conditions. (b) The QMC configuration of $\mathcal{Z}_{A}^{(2)}$. The configuration is two replicas with the sites in the entangling region $A$  glued together and sites in $\overline{A}$ with periodical boundary conditions. (c) The nonequilibrium increment process. We divide a nonequilibrium process which is characterized by $\lambda$ evolving from $\lambda=0$ to $\lambda=1$ into $K$ pieces. Each piece is independent with another and evolves from $\lambda=k\Delta$ to $\lambda=(k+1)\Delta$ with $k=0,1,...,K-1$ and $\Delta=1/K$. The final entanglement entropy is the summation of $K$ such independent nonequilibrium pieces.} 
\label{fig:fig2}
\end{figure}

{\it{Nonequilibrium increment method for entanglement entropy.-}}
Precise determination of the value of corner corrections, especially at QCPs, is by no mean an easy task.
 Below, we first introduce an improved estimator based on nonequilibrium increment method, which can substantially increase the precision and efficiency of the computation of R\'enyi EE in QMC simulations~\cite{JRZhao2021method}.

The $n$-th R\'enyi EE $S_{A}^{(n)}=\frac{\ln(\operatorname{Tr}\left(\rho_{\mathrm{A}}^{n}\right))}{1-n}$ at finite temperature can be re-expressed by the ratio of two partition functions
$S_{A}^{(n)}=\frac{1}{1-n} \ln \left(\frac{\mathcal{Z}_{A}^{(n)}}{\mathcal{Z}_{\varnothing}^{(n)}}\right)$, stemming from its trace structure in the path integral~\cite{Calabrese_2004}. Here A is the entangled region. In the configuration space of QMC simulation~\cite{Sandvik1999,Syljuaasen2002}, as shown in Fig.~\ref{fig:fig2}, $\mathcal{Z}_{A}^{(2)}$ is a partition function of two replicas with entangling region $A$ glued together and its complement $\overline{A}$ independent. $\mathcal{Z}_{\varnothing}^{(2)}$ can be viewed as a special case of  $\mathcal{Z}_{A}^{(2)}$ where $A$ is an empty set.

To evaluate $S_{A}^{(2)}$ on lattice models, various QMC estimators have been introduced~\cite{Hastings2010,Humeniuk2012,Inglis2013,Luitz2014,DEmidio2020}. While nearly all the algorithms suffer from the computational complexity -- simulating replicas of the space-time configuration and connecting them with different boundary conditions -- rendering poor quality data at low temperatures and large system sizes, the recent nonequilibrium measurement of the R\'enyi entanglement entropy~\cite{DEmidio2020} stands out for its reliability. 

However this method still has limitations (detailed analysis will be presented elsewhere~\cite{JRZhao2021method}). For large systems if the quench is not slow enough then not all sites in $A$ will join in the glued geometry as the end of the quench, leading to failure of the measurement. Although increasing the quench time can safely resolve this problem, it is often costly as the simulation time significantly increases with the quench time. Such limitations heavily affect the computation of EE on larger systems, especially the more complicated models beyond nearest-neigbhor Heisenberg.

Here we put forward an improved version of the nonequilibrium measurement -- the nonequilibrium increment method -- which can reduce the limitation of the nonequilibrium measurement on large systems. A schematic flow of the method is shown in Fig.~\ref{fig:fig2} (c) and it can be seen that our method divides a nonequilibrium process into many smaller paralleled processes, and these smaller processes can be computed independently and thus are ideal for highly parallel simulations, and in this way the nonequilibrium increment method decreases the simulation time as well as improves the data quality. We give a breif outline of our method below and will explain it in detail elsewhere~\cite{JRZhao2021method}.

In the nonequilibrium method~\cite{DEmidio2020}, one introduces a function $\mathcal{Z}_{A}^{(n)}(\lambda)$ which is the sum of a collection of partition functions $Z_{B}^{(n)}$ weighted by $g_{A}\left(\lambda, N_{B}\right)=\lambda^{N_{B}}(1-\lambda)^{N_{A}-N_{B}}$ where B is a subset of the entangled region A, $N_{A}$ is the total number of sites in A, and $N_{B}$ is the total number of sites in B, the R\'enyi EE can be expressed as an integral over $\lambda\in[0,1]$, $\mathcal{Z}_{A}^{(n)}(\lambda)=\sum_{B \subseteq A} g_{A}\left(\lambda, N_{B}\right) Z_{B}^{(n)}$
where $\mathcal{Z}_{A}^{(n)}(1)=\mathcal{Z}_{A}^{(n)}$ and $\mathcal{Z}_{A}^{(n)}(0)={\mathcal{Z}_{\varnothing}^{(n)}}$. Thus the entropy can be rewritten as $S_{A}^{(n)}=\frac{1}{1-n} \int_{0}^{1} d \lambda \frac{\partial \ln \mathcal{Z}_{A}^{(n)}(\lambda)}{\partial \lambda}$.
Ref.~\cite{DEmidio2020} puts forward a nonequilibrium work defined as
$W_{A}^{(n)}=-\frac{1}{\beta} \int_{t_{i}}^{t_{f}} d t \frac{d \lambda}{d t} \frac{\partial \ln g_{A}\left(\lambda(t), N_{B}(t)\right)}{\partial \lambda}$ with $\lambda(t_{i})=0$ and $\lambda(t_{f})=1$. According to the Jarzynski’s equality~\cite{Jarzynski1997} it can be proven that the entropy can be estimated by the work
\begin{equation}
S_{A}^{(n)}=\frac{1}{1-n} \ln \left(\left\langle e^{-\beta W_{A}^{(n)}}\right\rangle\right).
\end{equation}
even when the nonequilibrium process is at finite rate.
Our optimized method divides the integrating region $[0,1]$ into  many small regions e.g. $[0,\Delta],... [k\Delta,(k+1)\Delta],...,[1-\Delta,1]$, as shown in Fig.~\ref{fig:fig2} (c), then the nonequilibrium process can be seen as the sum of many small and independent nonequilibrium processes which can be simulated simultaneously. In this way, by means of massive parallel computing, we can greatly improve the data quality and reduce the simulation time.

\begin{figure*}[htp!]
\centering
\includegraphics[width=2\columnwidth]{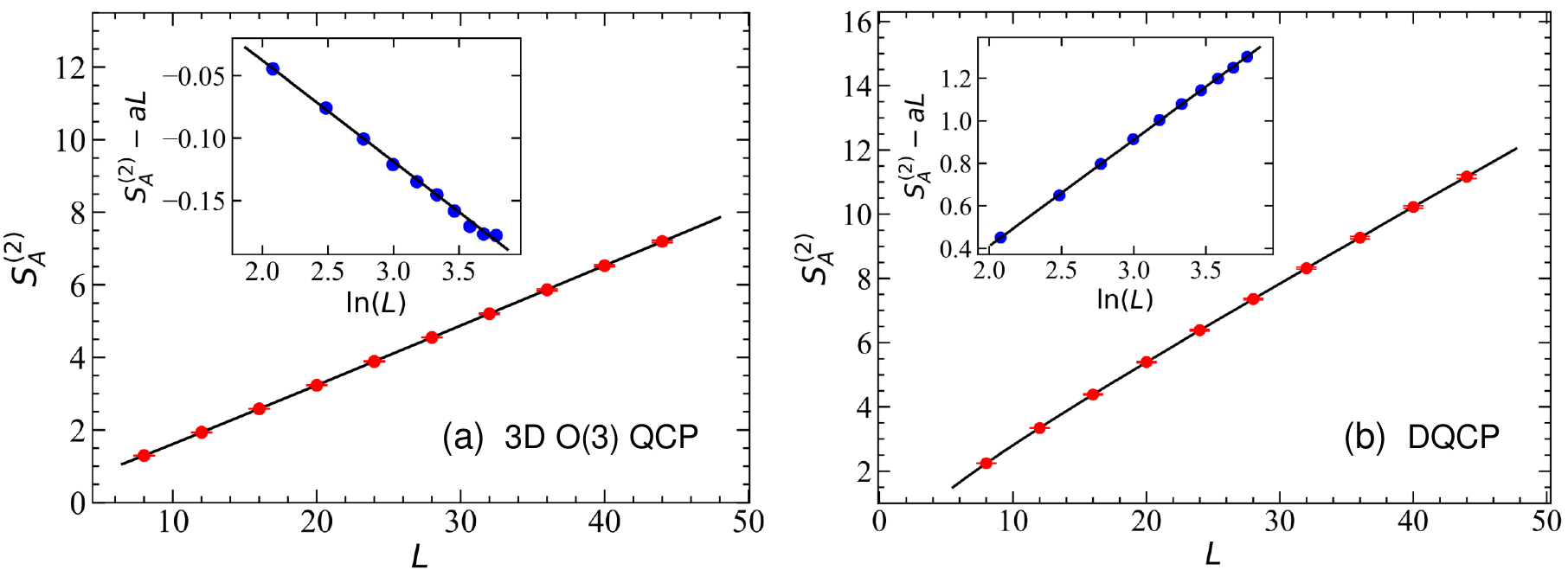}
\caption{$S^{(2)}_{A}(L)$ with even-B boundary of entangling region for (a) the $L\times L$ square lattice $H_{J_1-J_2}$ model at 3D O(3) QCP and (b) the $H_{J-Q_3}$ model at DQCP with pinning field $\delta =0.05$ . The fitting result for $H_{J_1-J_2}$ is  $S_{A}^{(2)}(L)=0.168(1)L-0.081(4)\ln(L)-0.124(7)$. The fitting result for $H_{J-Q_3}$ is  $S_{A}^{(2)}(L)=0.224(1)L+0.49(1)\ln(L)-0.58(2)$. Insets show the $S^{(2)}_A-aL$ versus $\ln(L)$ such that the sign of the log-corrections manifest. It is clear that the DQC log-correction accquires a opposite sign compared with the O(n) ones. This is in constrast with the positivity requirement of EE for unitary CFTs.}
\label{fig:fig3}
\end{figure*}

We start from the following formula
$\frac{\mathcal{Z}_{A}^{(n)}(1)}{\mathcal{Z}_{A}^{(n)}(0)}=\prod_{k=1}^{K}\frac{\mathcal{Z}_{A}^{(n)}(k\Delta)}{\mathcal{Z}_{A}^{(n)}((k-1)\Delta)}$, where $K$ is an integer and $\Delta=\frac{1}{K}$, and rewrite the R\'enyi entanglement entropy as $S_{A}^{(n)}=\frac{1}{1-n}\sum_{k=0,1,\cdots,K-1} \int_{k\Delta}^{(k+1)\Delta} d \lambda \frac{\partial \ln \mathcal{Z}_{A}^{(n)}(\lambda)}{\partial \lambda}$.
The argument for calculating the integral holds regardless of the lower and upper limits of the integral. As a result, we can apply the same argument on each individual integral and correspondingly get
\begin{equation}
    \label{eq:eq5}
	S_{A}^{(n)}=\frac{1}{1-n}\sum_{k=0,1,\cdots,K-1} \ln \left(\left\langle e^{-\beta W_{k,A}^{(n)}}\right\rangle\right)
\end{equation}
where $W_{k,A}^{(n)}$ is the work for the small piece between $\lambda(t_{i})=k\Delta$ and $\lambda(t_{f})=(k+1)\Delta$ in the nonequilibrium process. The detailed implementation protocol of our method is presented in SM~\cite{suppl}. We find such divide-and-conquer protocol give very robust results of the R\'enyi EE. And since now the nonequilibrium process are carried out in parallel in hundreds or thousands of short processes, the spendup with the same factor and the increasement of the data quality are self-evident.

{\it{Results.-}}
With the non-equilibrium increment method at hand, we apply it to measure the 2nd R\'enyi EE for the $H_{J_1-J_2}$ and $H_{J-Q_3}$ models, and reveal the scaling behavior of $S^{(2)}_A(L)$ at their corresponding QCPs.

The main results are shown in Fig.~\ref{fig:fig3}. Since the columnar dimers in $H_{J_1-J_2}$ break the lattice translation symmetry, we test three different types of the entangling region $A$, denoted as odd, even-A and even-B. Each type of regions gives a different value of the area-law coefficient since they cut different kinds of bonds, which also strongly affects the estimate of the subleading corner correction. We find the even-B regions, whose boundaries do not cut any dimer and thus have the smallest area-law coefficient, yield the most robust finite-size scaling behavior. We therefore adopt such a geometry for both $H_{J_1-J_2}$ and $H_{J-Q_3}$ presented here. Detailed discussions of the three different boundaries and their finite-size scaling behavior of $S^{(2)}_A(L)$ are presented in SM~\cite{suppl}.

The main panel of Fig.~\ref{fig:fig3} (a) shows finite size dependence of the $S^{(2)}_{A}(L)$ for the $H_{J_1-J_2}$ at its QCP. After fitting the data with Eq.~\eqref{eq:REE}, we obtain the coefficients $a$, $s$ and $b$. We find $s=0.081(4)$, very close to the prediction of Gaussion theory. In the inset, we plot the EE after extracting the area-law contribution, i.e. $S^{(2)}_{A}-aL$ versus $\ln(L)$. The linear dependence is quite clear with a negative slope, i.e. $s>0$. 

We now turn to the DQC of $H_{J-Q_3}$. Here the model is manifestly translation-invariant, so naively no special choice for the geometry of the entangling region is necessary. However, even at the DQC the VBS domains still exist in a finite-size system and can still cause uncertainty in the EE. To reduce such finite-size error, we employ a small pinning field to lock the VBS order to a fixed configuration, which allows us to work consistently with the even-B regions~\cite{Assaad2013,YCWang2021DQCdisorder}. The field is applied on the $J_2$ term such that $J_2=J+\frac{\delta}{L}$, i.e. when extrapolating to the thermodynamic limit, the simulated Hamiltonian goes back to the original $H_{J-Q_3}$. We find the value of $\delta=0.05$ gives the well-converged results of $S^{(2)}_{A}(L)$ and also note that even without the pinning field, $\delta=0$, the same qualitative conclusion, as in Fig.~\ref{fig:fig3} (b), still holds (see SM~\cite{suppl} for details where such universal behavior of the negative log-correction is persistently present from $\delta \in [0,0.15]$, representing the robustness of our observation.)


The fit of data in main panel of Fig.~\ref{fig:fig3} (b) accroding to Eq.~\eqref{eq:REE} gives rise to $s=-0.49(1)$.  After extracting the area law contribution, as shown in the inset of $S^{(2)}_{A}-aL$ versus $\ln L$, a straight line with positive slope, i.e. $s<0$ appears. Such large, negative value of $s$ (one magnitude larger than that in the O$(n)$ transition, the same large $s<0$ also holds even when $\delta=0$), is in sharp constrast to the expected corner corrections of the QCPs with CFT. And it is from here, our results unambiguously reveal fundamental differences between DQC and QCPs described by unitary CFTs. 

{\it{Discussions.-}} Our findings of the large and negative $s$ in the 2nd Renyi EE, raise a number of intriguing questions about the theory of DQC. Since a negative $s$ is not allowed in an unitary CFT, our observation appears to rule out such a description. This is consistent with recent conformal bootstrap study, which finds tension between bounds following from unitary conformal invariance with the numerically computed critical exponents~\cite{Shao2016,CWang2017,Nahum2015}. It is natural to connect these observations to the proposal that the observed regime of the DQC is controlled by a non-unitary fixed point very close to the physical parameter space~\cite{CWang2017, Nahum2020,RCMa2020,ZJWang2021}, which implies approximate conformal invariance within a large length scale. However, it is not clear whether such a scenario can naturally explain a relatively large and negative $s$. Since the complex fixed point has to be very close to the physical parameter space, one expects that the ``non-unitarity'', which manifests in e.g. imaginary part of scaling dimensions, should be quite small. This is indeed the case in known examples of weakly first-order transition controlled by a complex CFT  (note the suggestion that DQC is the precursors to a weakly first-order transition~\cite{Jiang2008,Kuklov2008,KChen2013,DEmidio2021}), such as the $Q=5$ Potts model in (1+1)d where the central charge, which is the coefficient of the log term in EE, appears to be a real positive number in numerical calculations~\cite{MaHe2019}. Our results however suggests that the violation of unitarity is not just a small complex correction. 

On the other hand, it is known that critical exponents of DQC exhibit unusual drift behavior with system size~\cite{Shao2016}.  It is possible that similar drift also occurs for $s$, and if it is the case the formula Eq. \eqref{eq:REE} needs to be corrected. In fact, a generic feature of theories controlled by complex CFTs is that various universal quantities, such as scaling dimensions, exhibit drifting (or walking RG in technical terms)~\cite{Gorbenko2018I,Gorbenko2018II}. It is therefore important to more systematically understand the finite-size correction to $s$ for complex CFTs, which we leave for future work. Other possible origins of the drift and how they affect the corner correction should also be investigated. For example, theoretically there exists a dangerously irrelevant operator at the critical point associated with the breaking of the emergent symmetry by lattice effect, which may introduce a new length scale in the problem. More recently, there are new evidence that shows the DQC is a multicritical point~\cite{BWZhao2020PRL}, it will also be interesting to investigate the scaling of R\'enyi EE in such modified models to verify the theoretical~\cite{DCLu2021} and numerical predictions~\cite{BWZhao2020PRL} therein.


{\it{Acknowledgements.-}} J.R.Z., Z.Y. and Z.Y.M. would like to thank Chenjie Wang for helpful discussions on the nonequilibrium process, they acknowledge support from the RGC of Hong Kong SAR of China (Grant Nos. 17303019, 17301420 and AoE/P-701/20), MOST through the National Key Research and Development Program (Grant No. 2016YFA0300502) and the Strategic Priority Research Program of the Chinese Academy of Sciences (Grant No. XDB33000000). Y.C.W. acknowledges the supports from the NSFC under Grant Nos.~11804383 and 11975024, the NSF of Jiangsu Province under Grant No.~BK20180637, and the Fundamental Research Funds for the Central Universities under Grant No. 2018QNA39. M.C. acknowledges support from NSF under award number DMR-1846109 and the Alfred P.~Sloan foundation. We thank the Computational Initiative at the Faculty of Science and the Information Technology Services at the University of Hong Kong and the Tianhe platforms at the National Supercomputer Centers in Tianjin and Guangzhou for their technical support and generous allocation of CPU time. The authors acknowledge Beijng PARATERA Tech CO.,Ltd. for providing HPC resources that have contributed to the research results reported within this paper. 

\bibliography{eedqcp}

\begin{thebibliography}{64}%
\makeatletter
\providecommand \@ifxundefined [1]{%
 \@ifx{#1\undefined}
}%
\providecommand \@ifnum [1]{%
 \ifnum #1\expandafter \@firstoftwo
 \else \expandafter \@secondoftwo
 \fi
}%
\providecommand \@ifx [1]{%
 \ifx #1\expandafter \@firstoftwo
 \else \expandafter \@secondoftwo
 \fi
}%
\providecommand \natexlab [1]{#1}%
\providecommand \enquote  [1]{``#1''}%
\providecommand \bibnamefont  [1]{#1}%
\providecommand \bibfnamefont [1]{#1}%
\providecommand \citenamefont [1]{#1}%
\providecommand \href@noop [0]{\@secondoftwo}%
\providecommand \href [0]{\begingroup \@sanitize@url \@href}%
\providecommand \@href[1]{\@@startlink{#1}\@@href}%
\providecommand \@@href[1]{\endgroup#1\@@endlink}%
\providecommand \@sanitize@url [0]{\catcode `\\12\catcode `\$12\catcode
  `\&12\catcode `\#12\catcode `\^12\catcode `\_12\catcode `\%12\relax}%
\providecommand \@@startlink[1]{}%
\providecommand \@@endlink[0]{}%
\providecommand \url  [0]{\begingroup\@sanitize@url \@url }%
\providecommand \@url [1]{\endgroup\@href {#1}{\urlprefix }}%
\providecommand \urlprefix  [0]{URL }%
\providecommand \Eprint [0]{\href }%
\providecommand \doibase [0]{http://dx.doi.org/}%
\providecommand \selectlanguage [0]{\@gobble}%
\providecommand \bibinfo  [0]{\@secondoftwo}%
\providecommand \bibfield  [0]{\@secondoftwo}%
\providecommand \translation [1]{[#1]}%
\providecommand \BibitemOpen [0]{}%
\providecommand \bibitemStop [0]{}%
\providecommand \bibitemNoStop [0]{.\EOS\space}%
\providecommand \EOS [0]{\spacefactor3000\relax}%
\providecommand \BibitemShut  [1]{\csname bibitem#1\endcsname}%
\let\auto@bib@innerbib\@empty
\bibitem [{\citenamefont {Calabrese}\ and\ \citenamefont
  {Cardy}(2004)}]{Calabrese_2004}%
  \BibitemOpen
  \bibfield  {author} {\bibinfo {author} {\bibfnamefont {P.}~\bibnamefont
  {Calabrese}}\ and\ \bibinfo {author} {\bibfnamefont {J.}~\bibnamefont
  {Cardy}},\ }\href {\doibase 10.1088/1742-5468/2004/06/p06002} {\bibfield
  {journal} {\bibinfo  {journal} {Journal of Statistical Mechanics: Theory and
  Experiment}\ }\textbf {\bibinfo {volume} {2004}},\ \bibinfo {pages} {P06002}
  (\bibinfo {year} {2004})}\BibitemShut {NoStop}%
\bibitem [{\citenamefont {Fradkin}\ and\ \citenamefont
  {Moore}(2006)}]{Fradkin2006}%
  \BibitemOpen
  \bibfield  {author} {\bibinfo {author} {\bibfnamefont {E.}~\bibnamefont
  {Fradkin}}\ and\ \bibinfo {author} {\bibfnamefont {J.~E.}\ \bibnamefont
  {Moore}},\ }\href {\doibase 10.1103/PhysRevLett.97.050404} {\bibfield
  {journal} {\bibinfo  {journal} {Phys. Rev. Lett.}\ }\textbf {\bibinfo
  {volume} {97}},\ \bibinfo {pages} {050404} (\bibinfo {year}
  {2006})}\BibitemShut {NoStop}%
\bibitem [{\citenamefont {Casini}\ and\ \citenamefont
  {Huerta}(2007{\natexlab{a}})}]{Casini2006}%
  \BibitemOpen
  \bibfield  {author} {\bibinfo {author} {\bibfnamefont {H.}~\bibnamefont
  {Casini}}\ and\ \bibinfo {author} {\bibfnamefont {M.}~\bibnamefont
  {Huerta}},\ }\href {\doibase 10.1016/j.nuclphysb.2006.12.012} {\bibfield
  {journal} {\bibinfo  {journal} {Nucl. Phys. B}\ }\textbf {\bibinfo {volume}
  {764}},\ \bibinfo {pages} {183} (\bibinfo {year} {2007}{\natexlab{a}})},\
  \Eprint {http://arxiv.org/abs/hep-th/0606256} {arXiv:hep-th/0606256}
  \BibitemShut {NoStop}%
\bibitem [{\citenamefont {Kitaev}\ and\ \citenamefont
  {Preskill}(2006)}]{Kitaev2006}%
  \BibitemOpen
  \bibfield  {author} {\bibinfo {author} {\bibfnamefont {A.}~\bibnamefont
  {Kitaev}}\ and\ \bibinfo {author} {\bibfnamefont {J.}~\bibnamefont
  {Preskill}},\ }\href {\doibase 10.1103/PhysRevLett.96.110404} {\bibfield
  {journal} {\bibinfo  {journal} {Phys. Rev. Lett.}\ }\textbf {\bibinfo
  {volume} {96}},\ \bibinfo {pages} {110404} (\bibinfo {year}
  {2006})}\BibitemShut {NoStop}%
\bibitem [{\citenamefont {Levin}\ and\ \citenamefont {Wen}(2006)}]{Levin2006}%
  \BibitemOpen
  \bibfield  {author} {\bibinfo {author} {\bibfnamefont {M.}~\bibnamefont
  {Levin}}\ and\ \bibinfo {author} {\bibfnamefont {X.-G.}\ \bibnamefont
  {Wen}},\ }\href {\doibase 10.1103/PhysRevLett.96.110405} {\bibfield
  {journal} {\bibinfo  {journal} {Phys. Rev. Lett.}\ }\textbf {\bibinfo
  {volume} {96}},\ \bibinfo {pages} {110405} (\bibinfo {year}
  {2006})}\BibitemShut {NoStop}%
\bibitem [{\citenamefont {Hastings}\ \emph
  {et~al.}(2010{\natexlab{a}})\citenamefont {Hastings}, \citenamefont
  {Gonz\'alez}, \citenamefont {Kallin},\ and\ \citenamefont
  {Melko}}]{Hastings2010}%
  \BibitemOpen
  \bibfield  {author} {\bibinfo {author} {\bibfnamefont {M.~B.}\ \bibnamefont
  {Hastings}}, \bibinfo {author} {\bibfnamefont {I.}~\bibnamefont
  {Gonz\'alez}}, \bibinfo {author} {\bibfnamefont {A.~B.}\ \bibnamefont
  {Kallin}}, \ and\ \bibinfo {author} {\bibfnamefont {R.~G.}\ \bibnamefont
  {Melko}},\ }\href {\doibase 10.1103/PhysRevLett.104.157201} {\bibfield
  {journal} {\bibinfo  {journal} {Phys. Rev. Lett.}\ }\textbf {\bibinfo
  {volume} {104}},\ \bibinfo {pages} {157201} (\bibinfo {year}
  {2010}{\natexlab{a}})}\BibitemShut {NoStop}%
\bibitem [{\citenamefont {{Metlitski}}\ and\ \citenamefont
  {{Grover}}(2011)}]{Metlitski2011}%
  \BibitemOpen
  \bibfield  {author} {\bibinfo {author} {\bibfnamefont {M.~A.}\ \bibnamefont
  {{Metlitski}}}\ and\ \bibinfo {author} {\bibfnamefont {T.}~\bibnamefont
  {{Grover}}},\ }\href@noop {} {\bibfield  {journal} {\bibinfo  {journal}
  {arXiv e-prints}\ ,\ \bibinfo {eid} {arXiv:1112.5166}} (\bibinfo {year}
  {2011})},\ \Eprint {http://arxiv.org/abs/1112.5166} {arXiv:1112.5166
  [cond-mat.str-el]} \BibitemShut {NoStop}%
\bibitem [{\citenamefont {Isakov}\ \emph {et~al.}(2011)\citenamefont {Isakov},
  \citenamefont {Hastings},\ and\ \citenamefont {Melko}}]{Isakov2011}%
  \BibitemOpen
  \bibfield  {author} {\bibinfo {author} {\bibfnamefont {S.~V.}\ \bibnamefont
  {Isakov}}, \bibinfo {author} {\bibfnamefont {M.~B.}\ \bibnamefont
  {Hastings}}, \ and\ \bibinfo {author} {\bibfnamefont {R.~G.}\ \bibnamefont
  {Melko}},\ }\href {\doibase 10.1038/nphys2036} {\bibfield  {journal}
  {\bibinfo  {journal} {Nature Physics}\ }\textbf {\bibinfo {volume} {7}},\
  \bibinfo {pages} {772 } (\bibinfo {year} {2011})}\BibitemShut {NoStop}%
\bibitem [{\citenamefont {Jiang}\ \emph {et~al.}(2012)\citenamefont {Jiang},
  \citenamefont {Wang},\ and\ \citenamefont {Balents}}]{Jiang2012}%
  \BibitemOpen
  \bibfield  {author} {\bibinfo {author} {\bibfnamefont {H.-C.}\ \bibnamefont
  {Jiang}}, \bibinfo {author} {\bibfnamefont {Z.}~\bibnamefont {Wang}}, \ and\
  \bibinfo {author} {\bibfnamefont {L.}~\bibnamefont {Balents}},\ }\href
  {\doibase 10.1038/nphys2465} {\bibfield  {journal} {\bibinfo  {journal}
  {Nature Physics}\ }\textbf {\bibinfo {volume} {8}},\ \bibinfo {pages} {902 }
  (\bibinfo {year} {2012})}\BibitemShut {NoStop}%
\bibitem [{\citenamefont {Casini}\ and\ \citenamefont
  {Huerta}(2012)}]{Casini2012}%
  \BibitemOpen
  \bibfield  {author} {\bibinfo {author} {\bibfnamefont {H.}~\bibnamefont
  {Casini}}\ and\ \bibinfo {author} {\bibfnamefont {M.}~\bibnamefont
  {Huerta}},\ }\href {\doibase 10.1007/JHEP11(2012)087} {\bibfield  {journal}
  {\bibinfo  {journal} {Journal of High Energy Physics}\ }\textbf {\bibinfo
  {volume} {2012}},\ \bibinfo {pages} {87} (\bibinfo {year}
  {2012})}\BibitemShut {NoStop}%
\bibitem [{\citenamefont {Swingle}\ and\ \citenamefont
  {Senthil}(2012)}]{Swingle2012}%
  \BibitemOpen
  \bibfield  {author} {\bibinfo {author} {\bibfnamefont {B.}~\bibnamefont
  {Swingle}}\ and\ \bibinfo {author} {\bibfnamefont {T.}~\bibnamefont
  {Senthil}},\ }\href {\doibase 10.1103/PhysRevB.86.155131} {\bibfield
  {journal} {\bibinfo  {journal} {Phys. Rev. B}\ }\textbf {\bibinfo {volume}
  {86}},\ \bibinfo {pages} {155131} (\bibinfo {year} {2012})}\BibitemShut
  {NoStop}%
\bibitem [{\citenamefont {Humeniuk}\ and\ \citenamefont
  {Roscilde}(2012{\natexlab{a}})}]{Humeniuk2012}%
  \BibitemOpen
  \bibfield  {author} {\bibinfo {author} {\bibfnamefont {S.}~\bibnamefont
  {Humeniuk}}\ and\ \bibinfo {author} {\bibfnamefont {T.}~\bibnamefont
  {Roscilde}},\ }\href {\doibase 10.1103/PhysRevB.86.235116} {\bibfield
  {journal} {\bibinfo  {journal} {Phys. Rev. B}\ }\textbf {\bibinfo {volume}
  {86}},\ \bibinfo {pages} {235116} (\bibinfo {year}
  {2012}{\natexlab{a}})}\BibitemShut {NoStop}%
\bibitem [{\citenamefont {Inglis}\ and\ \citenamefont
  {Melko}(2013)}]{Inglis2013}%
  \BibitemOpen
  \bibfield  {author} {\bibinfo {author} {\bibfnamefont {S.}~\bibnamefont
  {Inglis}}\ and\ \bibinfo {author} {\bibfnamefont {R.~G.}\ \bibnamefont
  {Melko}},\ }\href {\doibase 10.1103/PhysRevE.87.013306} {\bibfield  {journal}
  {\bibinfo  {journal} {Phys. Rev. E}\ }\textbf {\bibinfo {volume} {87}},\
  \bibinfo {pages} {013306} (\bibinfo {year} {2013})}\BibitemShut {NoStop}%
\bibitem [{\citenamefont {{Inglis}}\ and\ \citenamefont
  {{Melko}}(2013)}]{InglisNJP2013}%
  \BibitemOpen
  \bibfield  {author} {\bibinfo {author} {\bibfnamefont {S.}~\bibnamefont
  {{Inglis}}}\ and\ \bibinfo {author} {\bibfnamefont {R.~G.}\ \bibnamefont
  {{Melko}}},\ }\href {\doibase 10.1088/1367-2630/15/7/073048} {\bibfield
  {journal} {\bibinfo  {journal} {New J. Phys}\ }\textbf {\bibinfo {volume}
  {15}},\ \bibinfo {pages} {073048} (\bibinfo {year} {2013})},\ \Eprint
  {http://arxiv.org/abs/1305.1069} {arXiv:1305.1069} \BibitemShut {NoStop}%
\bibitem [{\citenamefont {Kallin}\ \emph {et~al.}(2013)\citenamefont {Kallin},
  \citenamefont {Hyatt}, \citenamefont {Singh},\ and\ \citenamefont
  {Melko}}]{KallinPRL2013}%
  \BibitemOpen
  \bibfield  {author} {\bibinfo {author} {\bibfnamefont {A.~B.}\ \bibnamefont
  {Kallin}}, \bibinfo {author} {\bibfnamefont {K.}~\bibnamefont {Hyatt}},
  \bibinfo {author} {\bibfnamefont {R.~R.~P.}\ \bibnamefont {Singh}}, \ and\
  \bibinfo {author} {\bibfnamefont {R.~G.}\ \bibnamefont {Melko}},\ }\href
  {\doibase 10.1103/PhysRevLett.110.135702} {\bibfield  {journal} {\bibinfo
  {journal} {Phys. Rev. Lett.}\ }\textbf {\bibinfo {volume} {110}},\ \bibinfo
  {pages} {135702} (\bibinfo {year} {2013})}\BibitemShut {NoStop}%
\bibitem [{\citenamefont {Luitz}\ \emph {et~al.}(2014)\citenamefont {Luitz},
  \citenamefont {Plat}, \citenamefont {Laflorencie},\ and\ \citenamefont
  {Alet}}]{Luitz2014}%
  \BibitemOpen
  \bibfield  {author} {\bibinfo {author} {\bibfnamefont {D.~J.}\ \bibnamefont
  {Luitz}}, \bibinfo {author} {\bibfnamefont {X.}~\bibnamefont {Plat}},
  \bibinfo {author} {\bibfnamefont {N.}~\bibnamefont {Laflorencie}}, \ and\
  \bibinfo {author} {\bibfnamefont {F.}~\bibnamefont {Alet}},\ }\href {\doibase
  10.1103/PhysRevB.90.125105} {\bibfield  {journal} {\bibinfo  {journal} {Phys.
  Rev. B}\ }\textbf {\bibinfo {volume} {90}},\ \bibinfo {pages} {125105}
  (\bibinfo {year} {2014})}\BibitemShut {NoStop}%
\bibitem [{\citenamefont {{Kallin}}\ \emph {et~al.}(2014)\citenamefont
  {{Kallin}}, \citenamefont {{Stoudenmire}}, \citenamefont {{Fendley}},
  \citenamefont {{Singh}},\ and\ \citenamefont {{Melko}}}]{KallinJS2014}%
  \BibitemOpen
  \bibfield  {author} {\bibinfo {author} {\bibfnamefont {A.~B.}\ \bibnamefont
  {{Kallin}}}, \bibinfo {author} {\bibfnamefont {E.~M.}\ \bibnamefont
  {{Stoudenmire}}}, \bibinfo {author} {\bibfnamefont {P.}~\bibnamefont
  {{Fendley}}}, \bibinfo {author} {\bibfnamefont {R.~R.~P.}\ \bibnamefont
  {{Singh}}}, \ and\ \bibinfo {author} {\bibfnamefont {R.~G.}\ \bibnamefont
  {{Melko}}},\ }\href {\doibase 10.1088/1742-5468/2014/06/P06009} {\bibfield
  {journal} {\bibinfo  {journal} {J. Stat. Mech.}\ }\textbf {\bibinfo {volume}
  {2014}},\ \bibinfo {eid} {06009} (\bibinfo {year} {2014})},\ \Eprint
  {http://arxiv.org/abs/1401.3504} {arXiv:1401.3504} \BibitemShut {NoStop}%
\bibitem [{\citenamefont {Helmes}\ and\ \citenamefont
  {Wessel}(2014)}]{Helmes2014}%
  \BibitemOpen
  \bibfield  {author} {\bibinfo {author} {\bibfnamefont {J.}~\bibnamefont
  {Helmes}}\ and\ \bibinfo {author} {\bibfnamefont {S.}~\bibnamefont
  {Wessel}},\ }\href {\doibase 10.1103/PhysRevB.89.245120} {\bibfield
  {journal} {\bibinfo  {journal} {Phys. Rev. B}\ }\textbf {\bibinfo {volume}
  {89}},\ \bibinfo {pages} {245120} (\bibinfo {year} {2014})}\BibitemShut
  {NoStop}%
\bibitem [{\citenamefont {Laflorencie}(2016)}]{Laflorencie2016}%
  \BibitemOpen
  \bibfield  {author} {\bibinfo {author} {\bibfnamefont {N.}~\bibnamefont
  {Laflorencie}},\ }\href {\doibase
  https://doi.org/10.1016/j.physrep.2016.06.008} {\bibfield  {journal}
  {\bibinfo  {journal} {Physics Reports}\ }\textbf {\bibinfo {volume} {646}},\
  \bibinfo {pages} {1} (\bibinfo {year} {2016})},\ \bibinfo {note} {quantum
  entanglement in condensed matter systems}\BibitemShut {NoStop}%
\bibitem [{\citenamefont {Wolf}(2006)}]{PhysRevLett.96.010404}%
  \BibitemOpen
  \bibfield  {author} {\bibinfo {author} {\bibfnamefont {M.~M.}\ \bibnamefont
  {Wolf}},\ }\href {\doibase 10.1103/PhysRevLett.96.010404} {\bibfield
  {journal} {\bibinfo  {journal} {Phys. Rev. Lett.}\ }\textbf {\bibinfo
  {volume} {96}},\ \bibinfo {pages} {010404} (\bibinfo {year}
  {2006})}\BibitemShut {NoStop}%
\bibitem [{\citenamefont {Lin}\ \emph {et~al.}(2007)\citenamefont {Lin},
  \citenamefont {Igl\'oi},\ and\ \citenamefont
  {Rieger}}]{PhysRevLett.99.147202}%
  \BibitemOpen
  \bibfield  {author} {\bibinfo {author} {\bibfnamefont {Y.-C.}\ \bibnamefont
  {Lin}}, \bibinfo {author} {\bibfnamefont {F.}~\bibnamefont {Igl\'oi}}, \ and\
  \bibinfo {author} {\bibfnamefont {H.}~\bibnamefont {Rieger}},\ }\href
  {\doibase 10.1103/PhysRevLett.99.147202} {\bibfield  {journal} {\bibinfo
  {journal} {Phys. Rev. Lett.}\ }\textbf {\bibinfo {volume} {99}},\ \bibinfo
  {pages} {147202} (\bibinfo {year} {2007})}\BibitemShut {NoStop}%
\bibitem [{\citenamefont {Yu}\ \emph {et~al.}(2008)\citenamefont {Yu},
  \citenamefont {Saleur},\ and\ \citenamefont {Haas}}]{PhysRevB.77.140402}%
  \BibitemOpen
  \bibfield  {author} {\bibinfo {author} {\bibfnamefont {R.}~\bibnamefont
  {Yu}}, \bibinfo {author} {\bibfnamefont {H.}~\bibnamefont {Saleur}}, \ and\
  \bibinfo {author} {\bibfnamefont {S.}~\bibnamefont {Haas}},\ }\href {\doibase
  10.1103/PhysRevB.77.140402} {\bibfield  {journal} {\bibinfo  {journal} {Phys.
  Rev. B}\ }\textbf {\bibinfo {volume} {77}},\ \bibinfo {pages} {140402}
  (\bibinfo {year} {2008})}\BibitemShut {NoStop}%
\bibitem [{\citenamefont {Kov\'acs}\ \emph {et~al.}(2012)\citenamefont
  {Kov\'acs}, \citenamefont {Igl\'oi},\ and\ \citenamefont
  {Cardy}}]{PhysRevB.86.214203}%
  \BibitemOpen
  \bibfield  {author} {\bibinfo {author} {\bibfnamefont {I.~A.}\ \bibnamefont
  {Kov\'acs}}, \bibinfo {author} {\bibfnamefont {F.}~\bibnamefont {Igl\'oi}}, \
  and\ \bibinfo {author} {\bibfnamefont {J.}~\bibnamefont {Cardy}},\ }\href
  {\doibase 10.1103/PhysRevB.86.214203} {\bibfield  {journal} {\bibinfo
  {journal} {Phys. Rev. B}\ }\textbf {\bibinfo {volume} {86}},\ \bibinfo
  {pages} {214203} (\bibinfo {year} {2012})}\BibitemShut {NoStop}%
\bibitem [{\citenamefont {Kov\'acs}\ and\ \citenamefont
  {Igl\'oi}(2012)}]{ref12}%
  \BibitemOpen
  \bibfield  {author} {\bibinfo {author} {\bibfnamefont {I.~A.}\ \bibnamefont
  {Kov\'acs}}\ and\ \bibinfo {author} {\bibfnamefont {F.}~\bibnamefont
  {Igl\'oi}},\ }\href {\doibase 10.1209/0295-5075/97/67009} {\bibfield
  {journal} {\bibinfo  {journal} {EPL}\ }\textbf {\bibinfo {volume} {97}},\
  \bibinfo {pages} {67009} (\bibinfo {year} {2012})}\BibitemShut {NoStop}%
\bibitem [{\citenamefont {{Zhao}}\ \emph {et~al.}(2020)\citenamefont {{Zhao}},
  \citenamefont {{Yan}}, \citenamefont {{Cheng}},\ and\ \citenamefont
  {{Meng}}}]{JRZhao2020}%
  \BibitemOpen
  \bibfield  {author} {\bibinfo {author} {\bibfnamefont {J.}~\bibnamefont
  {{Zhao}}}, \bibinfo {author} {\bibfnamefont {Z.}~\bibnamefont {{Yan}}},
  \bibinfo {author} {\bibfnamefont {M.}~\bibnamefont {{Cheng}}}, \ and\
  \bibinfo {author} {\bibfnamefont {Z.~Y.}\ \bibnamefont {{Meng}}},\
  }\href@noop {} {\bibfield  {journal} {\bibinfo  {journal} {arXiv e-prints}\ }
  (\bibinfo {year} {2020})},\ \Eprint {http://arxiv.org/abs/2011.12543}
  {arXiv:2011.12543} \BibitemShut {NoStop}%
\bibitem [{\citenamefont {Sandvik}(2007)}]{Sandvik2007}%
  \BibitemOpen
  \bibfield  {author} {\bibinfo {author} {\bibfnamefont {A.~W.}\ \bibnamefont
  {Sandvik}},\ }\href {\doibase 10.1103/PhysRevLett.98.227202} {\bibfield
  {journal} {\bibinfo  {journal} {Phys. Rev. Lett.}\ }\textbf {\bibinfo
  {volume} {98}},\ \bibinfo {pages} {227202} (\bibinfo {year}
  {2007})}\BibitemShut {NoStop}%
\bibitem [{\citenamefont {Senthil}\ \emph {et~al.}(2004)\citenamefont
  {Senthil}, \citenamefont {Balents}, \citenamefont {Sachdev}, \citenamefont
  {Vishwanath},\ and\ \citenamefont {Fisher}}]{Senthil2004}%
  \BibitemOpen
  \bibfield  {author} {\bibinfo {author} {\bibfnamefont {T.}~\bibnamefont
  {Senthil}}, \bibinfo {author} {\bibfnamefont {L.}~\bibnamefont {Balents}},
  \bibinfo {author} {\bibfnamefont {S.}~\bibnamefont {Sachdev}}, \bibinfo
  {author} {\bibfnamefont {A.}~\bibnamefont {Vishwanath}}, \ and\ \bibinfo
  {author} {\bibfnamefont {M.~P.~A.}\ \bibnamefont {Fisher}},\ }\href {\doibase
  10.1103/PhysRevB.70.144407} {\bibfield  {journal} {\bibinfo  {journal} {Phys.
  Rev. B}\ }\textbf {\bibinfo {volume} {70}},\ \bibinfo {pages} {144407}
  (\bibinfo {year} {2004})}\BibitemShut {NoStop}%
\bibitem [{\citenamefont {Ma}\ \emph {et~al.}(2018{\natexlab{a}})\citenamefont
  {Ma}, \citenamefont {Sun}, \citenamefont {You}, \citenamefont {Xu},
  \citenamefont {Vishwanath}, \citenamefont {Sandvik},\ and\ \citenamefont
  {Meng}}]{NSMa2018}%
  \BibitemOpen
  \bibfield  {author} {\bibinfo {author} {\bibfnamefont {N.}~\bibnamefont
  {Ma}}, \bibinfo {author} {\bibfnamefont {G.-Y.}\ \bibnamefont {Sun}},
  \bibinfo {author} {\bibfnamefont {Y.-Z.}\ \bibnamefont {You}}, \bibinfo
  {author} {\bibfnamefont {C.}~\bibnamefont {Xu}}, \bibinfo {author}
  {\bibfnamefont {A.}~\bibnamefont {Vishwanath}}, \bibinfo {author}
  {\bibfnamefont {A.~W.}\ \bibnamefont {Sandvik}}, \ and\ \bibinfo {author}
  {\bibfnamefont {Z.~Y.}\ \bibnamefont {Meng}},\ }\href {\doibase
  10.1103/PhysRevB.98.174421} {\bibfield  {journal} {\bibinfo  {journal} {Phys.
  Rev. B}\ }\textbf {\bibinfo {volume} {98}},\ \bibinfo {pages} {174421}
  (\bibinfo {year} {2018}{\natexlab{a}})}\BibitemShut {NoStop}%
\bibitem [{\citenamefont {Liu}\ \emph {et~al.}(2019)\citenamefont {Liu},
  \citenamefont {Wang}, \citenamefont {Sato}, \citenamefont {Hohenadler},
  \citenamefont {Wang}, \citenamefont {Guo},\ and\ \citenamefont
  {Assaad}}]{Liu_2019}%
  \BibitemOpen
  \bibfield  {author} {\bibinfo {author} {\bibfnamefont {Y.}~\bibnamefont
  {Liu}}, \bibinfo {author} {\bibfnamefont {Z.}~\bibnamefont {Wang}}, \bibinfo
  {author} {\bibfnamefont {T.}~\bibnamefont {Sato}}, \bibinfo {author}
  {\bibfnamefont {M.}~\bibnamefont {Hohenadler}}, \bibinfo {author}
  {\bibfnamefont {C.}~\bibnamefont {Wang}}, \bibinfo {author} {\bibfnamefont
  {W.}~\bibnamefont {Guo}}, \ and\ \bibinfo {author} {\bibfnamefont {F.~F.}\
  \bibnamefont {Assaad}},\ }\href {\doibase 10.1038/s41467-019-10372-0}
  {\bibfield  {journal} {\bibinfo  {journal} {Nat. Commun.}\ }\textbf {\bibinfo
  {volume} {10}} (\bibinfo {year} {2019}),\
  10.1038/s41467-019-10372-0}\BibitemShut {NoStop}%
\bibitem [{\citenamefont {Jiang}\ \emph {et~al.}(2008)\citenamefont {Jiang},
  \citenamefont {Nyfeler}, \citenamefont {Chandrasekharan},\ and\ \citenamefont
  {Wiese}}]{Jiang2008}%
  \BibitemOpen
  \bibfield  {author} {\bibinfo {author} {\bibfnamefont {F.-J.}\ \bibnamefont
  {Jiang}}, \bibinfo {author} {\bibfnamefont {M.}~\bibnamefont {Nyfeler}},
  \bibinfo {author} {\bibfnamefont {S.}~\bibnamefont {Chandrasekharan}}, \ and\
  \bibinfo {author} {\bibfnamefont {U.-J.}\ \bibnamefont {Wiese}},\ }\href
  {\doibase 10.1088/1742-5468/2008/02/p02009} {\bibfield  {journal} {\bibinfo
  {journal} {Journal of Statistical Mechanics: Theory and Experiment}\ }\textbf
  {\bibinfo {volume} {2008}},\ \bibinfo {pages} {P02009} (\bibinfo {year}
  {2008})}\BibitemShut {NoStop}%
\bibitem [{\citenamefont {Banerjee}\ \emph {et~al.}(2013)\citenamefont
  {Banerjee}, \citenamefont {Jiang}, \citenamefont {Widmer},\ and\
  \citenamefont {Wiese}}]{Banerjee:2013dda}%
  \BibitemOpen
  \bibfield  {author} {\bibinfo {author} {\bibfnamefont {D.}~\bibnamefont
  {Banerjee}}, \bibinfo {author} {\bibfnamefont {F.~J.}\ \bibnamefont {Jiang}},
  \bibinfo {author} {\bibfnamefont {P.}~\bibnamefont {Widmer}}, \ and\ \bibinfo
  {author} {\bibfnamefont {U.~J.}\ \bibnamefont {Wiese}},\ }\href {\doibase
  10.1088/1742-5468/2013/12/P12010} {\bibfield  {journal} {\bibinfo  {journal}
  {J. Stat. Mech.}\ }\textbf {\bibinfo {volume} {1312}},\ \bibinfo {pages}
  {P12010} (\bibinfo {year} {2013})},\ \Eprint {http://arxiv.org/abs/1303.6858}
  {arXiv:1303.6858 [cond-mat.str-el]} \BibitemShut {NoStop}%
\bibitem [{\citenamefont {Wang}\ \emph {et~al.}(2017)\citenamefont {Wang},
  \citenamefont {Nahum}, \citenamefont {Metlitski}, \citenamefont {Xu},\ and\
  \citenamefont {Senthil}}]{CWang2017}%
  \BibitemOpen
  \bibfield  {author} {\bibinfo {author} {\bibfnamefont {C.}~\bibnamefont
  {Wang}}, \bibinfo {author} {\bibfnamefont {A.}~\bibnamefont {Nahum}},
  \bibinfo {author} {\bibfnamefont {M.~A.}\ \bibnamefont {Metlitski}}, \bibinfo
  {author} {\bibfnamefont {C.}~\bibnamefont {Xu}}, \ and\ \bibinfo {author}
  {\bibfnamefont {T.}~\bibnamefont {Senthil}},\ }\href {\doibase
  10.1103/PhysRevX.7.031051} {\bibfield  {journal} {\bibinfo  {journal} {Phys.
  Rev. X}\ }\textbf {\bibinfo {volume} {7}},\ \bibinfo {pages} {031051}
  (\bibinfo {year} {2017})}\BibitemShut {NoStop}%
\bibitem [{\citenamefont {{Lu}}\ \emph {et~al.}(2021)\citenamefont {{Lu}},
  \citenamefont {{Xu}},\ and\ \citenamefont {{You}}}]{DCLu2021}%
  \BibitemOpen
  \bibfield  {author} {\bibinfo {author} {\bibfnamefont {D.-C.}\ \bibnamefont
  {{Lu}}}, \bibinfo {author} {\bibfnamefont {C.}~\bibnamefont {{Xu}}}, \ and\
  \bibinfo {author} {\bibfnamefont {Y.-Z.}\ \bibnamefont {{You}}},\ }\href@noop
  {} {\bibfield  {journal} {\bibinfo  {journal} {arXiv e-prints}\ ,\ \bibinfo
  {eid} {arXiv:2104.05147}} (\bibinfo {year} {2021})},\ \Eprint
  {http://arxiv.org/abs/2104.05147} {arXiv:2104.05147 [cond-mat.str-el]}
  \BibitemShut {NoStop}%
\bibitem [{\citenamefont {Lou}\ \emph {et~al.}(2009)\citenamefont {Lou},
  \citenamefont {Sandvik},\ and\ \citenamefont {Kawashima}}]{Lou2009}%
  \BibitemOpen
  \bibfield  {author} {\bibinfo {author} {\bibfnamefont {J.}~\bibnamefont
  {Lou}}, \bibinfo {author} {\bibfnamefont {A.~W.}\ \bibnamefont {Sandvik}}, \
  and\ \bibinfo {author} {\bibfnamefont {N.}~\bibnamefont {Kawashima}},\ }\href
  {\doibase 10.1103/PhysRevB.80.180414} {\bibfield  {journal} {\bibinfo
  {journal} {Phys. Rev. B}\ }\textbf {\bibinfo {volume} {80}},\ \bibinfo
  {pages} {180414} (\bibinfo {year} {2009})}\BibitemShut {NoStop}%
\bibitem [{\citenamefont {Ma}\ \emph {et~al.}(2018{\natexlab{b}})\citenamefont
  {Ma}, \citenamefont {Weinberg}, \citenamefont {Shao}, \citenamefont {Guo},
  \citenamefont {Yao},\ and\ \citenamefont {Sandvik}}]{NSMa2018anomalous}%
  \BibitemOpen
  \bibfield  {author} {\bibinfo {author} {\bibfnamefont {N.}~\bibnamefont
  {Ma}}, \bibinfo {author} {\bibfnamefont {P.}~\bibnamefont {Weinberg}},
  \bibinfo {author} {\bibfnamefont {H.}~\bibnamefont {Shao}}, \bibinfo {author}
  {\bibfnamefont {W.}~\bibnamefont {Guo}}, \bibinfo {author} {\bibfnamefont
  {D.-X.}\ \bibnamefont {Yao}}, \ and\ \bibinfo {author} {\bibfnamefont
  {A.~W.}\ \bibnamefont {Sandvik}},\ }\href {\doibase
  10.1103/PhysRevLett.121.117202} {\bibfield  {journal} {\bibinfo  {journal}
  {Phys. Rev. Lett.}\ }\textbf {\bibinfo {volume} {121}},\ \bibinfo {pages}
  {117202} (\bibinfo {year} {2018}{\natexlab{b}})}\BibitemShut {NoStop}%
\bibitem [{\citenamefont {Grover}(2013)}]{Grover2013}%
  \BibitemOpen
  \bibfield  {author} {\bibinfo {author} {\bibfnamefont {T.}~\bibnamefont
  {Grover}},\ }\href {\doibase 10.1103/PhysRevLett.111.130402} {\bibfield
  {journal} {\bibinfo  {journal} {Phys. Rev. Lett.}\ }\textbf {\bibinfo
  {volume} {111}},\ \bibinfo {pages} {130402} (\bibinfo {year}
  {2013})}\BibitemShut {NoStop}%
\bibitem [{\citenamefont {Assaad}(2015)}]{Assaad2015}%
  \BibitemOpen
  \bibfield  {author} {\bibinfo {author} {\bibfnamefont {F.~F.}\ \bibnamefont
  {Assaad}},\ }\href {\doibase 10.1103/PhysRevB.91.125146} {\bibfield
  {journal} {\bibinfo  {journal} {Phys. Rev. B}\ }\textbf {\bibinfo {volume}
  {91}},\ \bibinfo {pages} {125146} (\bibinfo {year} {2015})}\BibitemShut
  {NoStop}%
\bibitem [{\citenamefont {D'Emidio}(2020)}]{DEmidio2020}%
  \BibitemOpen
  \bibfield  {author} {\bibinfo {author} {\bibfnamefont {J.}~\bibnamefont
  {D'Emidio}},\ }\href {\doibase 10.1103/PhysRevLett.124.110602} {\bibfield
  {journal} {\bibinfo  {journal} {Phys. Rev. Lett.}\ }\textbf {\bibinfo
  {volume} {124}},\ \bibinfo {pages} {110602} (\bibinfo {year}
  {2020})}\BibitemShut {NoStop}%
\bibitem [{\citenamefont {Hastings}\ \emph
  {et~al.}(2010{\natexlab{b}})\citenamefont {Hastings}, \citenamefont
  {Gonz\'alez}, \citenamefont {Kallin},\ and\ \citenamefont
  {Melko}}]{PhysRevLett.104.157201}%
  \BibitemOpen
  \bibfield  {author} {\bibinfo {author} {\bibfnamefont {M.~B.}\ \bibnamefont
  {Hastings}}, \bibinfo {author} {\bibfnamefont {I.}~\bibnamefont
  {Gonz\'alez}}, \bibinfo {author} {\bibfnamefont {A.~B.}\ \bibnamefont
  {Kallin}}, \ and\ \bibinfo {author} {\bibfnamefont {R.~G.}\ \bibnamefont
  {Melko}},\ }\href {\doibase 10.1103/PhysRevLett.104.157201} {\bibfield
  {journal} {\bibinfo  {journal} {Phys. Rev. Lett.}\ }\textbf {\bibinfo
  {volume} {104}},\ \bibinfo {pages} {157201} (\bibinfo {year}
  {2010}{\natexlab{b}})}\BibitemShut {NoStop}%
\bibitem [{\citenamefont {Song}\ \emph {et~al.}(2011)\citenamefont {Song},
  \citenamefont {Laflorencie}, \citenamefont {Rachel},\ and\ \citenamefont
  {Le~Hur}}]{PhysRevB.83.224410}%
  \BibitemOpen
  \bibfield  {author} {\bibinfo {author} {\bibfnamefont {H.~F.}\ \bibnamefont
  {Song}}, \bibinfo {author} {\bibfnamefont {N.}~\bibnamefont {Laflorencie}},
  \bibinfo {author} {\bibfnamefont {S.}~\bibnamefont {Rachel}}, \ and\ \bibinfo
  {author} {\bibfnamefont {K.}~\bibnamefont {Le~Hur}},\ }\href {\doibase
  10.1103/PhysRevB.83.224410} {\bibfield  {journal} {\bibinfo  {journal} {Phys.
  Rev. B}\ }\textbf {\bibinfo {volume} {83}},\ \bibinfo {pages} {224410}
  (\bibinfo {year} {2011})}\BibitemShut {NoStop}%
\bibitem [{\citenamefont {Kallin}\ \emph {et~al.}(2011)\citenamefont {Kallin},
  \citenamefont {Hastings}, \citenamefont {Melko},\ and\ \citenamefont
  {Singh}}]{PhysRevB.84.165134}%
  \BibitemOpen
  \bibfield  {author} {\bibinfo {author} {\bibfnamefont {A.~B.}\ \bibnamefont
  {Kallin}}, \bibinfo {author} {\bibfnamefont {M.~B.}\ \bibnamefont
  {Hastings}}, \bibinfo {author} {\bibfnamefont {R.~G.}\ \bibnamefont {Melko}},
  \ and\ \bibinfo {author} {\bibfnamefont {R.~R.~P.}\ \bibnamefont {Singh}},\
  }\href {\doibase 10.1103/PhysRevB.84.165134} {\bibfield  {journal} {\bibinfo
  {journal} {Phys. Rev. B}\ }\textbf {\bibinfo {volume} {84}},\ \bibinfo
  {pages} {165134} (\bibinfo {year} {2011})}\BibitemShut {NoStop}%
\bibitem [{\citenamefont {Humeniuk}\ and\ \citenamefont
  {Roscilde}(2012{\natexlab{b}})}]{PhysRevB.86.235116}%
  \BibitemOpen
  \bibfield  {author} {\bibinfo {author} {\bibfnamefont {S.}~\bibnamefont
  {Humeniuk}}\ and\ \bibinfo {author} {\bibfnamefont {T.}~\bibnamefont
  {Roscilde}},\ }\href {\doibase 10.1103/PhysRevB.86.235116} {\bibfield
  {journal} {\bibinfo  {journal} {Phys. Rev. B}\ }\textbf {\bibinfo {volume}
  {86}},\ \bibinfo {pages} {235116} (\bibinfo {year}
  {2012}{\natexlab{b}})}\BibitemShut {NoStop}%
\bibitem [{\citenamefont {Wang}\ and\ \citenamefont
  {Davis}(2020)}]{PhysRevA.102.062413}%
  \BibitemOpen
  \bibfield  {author} {\bibinfo {author} {\bibfnamefont {Z.}~\bibnamefont
  {Wang}}\ and\ \bibinfo {author} {\bibfnamefont {E.~J.}\ \bibnamefont
  {Davis}},\ }\href {\doibase 10.1103/PhysRevA.102.062413} {\bibfield
  {journal} {\bibinfo  {journal} {Phys. Rev. A}\ }\textbf {\bibinfo {volume}
  {102}},\ \bibinfo {pages} {062413} (\bibinfo {year} {2020})}\BibitemShut
  {NoStop}%
\bibitem [{\citenamefont {Zhao}\ and\ \citenamefont {et.
  al.}(2021)}]{JRZhao2021method}%
  \BibitemOpen
  \bibfield  {author} {\bibinfo {author} {\bibfnamefont {J.}~\bibnamefont
  {Zhao}}\ and\ \bibinfo {author} {\bibnamefont {et. al.}},\ }\href@noop {}
  {\bibfield  {journal} {\bibinfo  {journal} {In preparation.}\ } (\bibinfo
  {year} {2021})}\BibitemShut {NoStop}%
\bibitem [{\citenamefont {Casini}\ and\ \citenamefont
  {Huerta}(2007{\natexlab{b}})}]{CASINI2007183}%
  \BibitemOpen
  \bibfield  {author} {\bibinfo {author} {\bibfnamefont {H.}~\bibnamefont
  {Casini}}\ and\ \bibinfo {author} {\bibfnamefont {M.}~\bibnamefont
  {Huerta}},\ }\href {\doibase https://doi.org/10.1016/j.nuclphysb.2006.12.012}
  {\bibfield  {journal} {\bibinfo  {journal} {Nuclear Physics B}\ }\textbf
  {\bibinfo {volume} {764}},\ \bibinfo {pages} {183} (\bibinfo {year}
  {2007}{\natexlab{b}})}\BibitemShut {NoStop}%
\bibitem [{\citenamefont {Bueno}\ \emph {et~al.}(2015)\citenamefont {Bueno},
  \citenamefont {Myers},\ and\ \citenamefont {Witczak-Krempa}}]{Bueno_2015}%
  \BibitemOpen
  \bibfield  {author} {\bibinfo {author} {\bibfnamefont {P.}~\bibnamefont
  {Bueno}}, \bibinfo {author} {\bibfnamefont {R.~C.}\ \bibnamefont {Myers}}, \
  and\ \bibinfo {author} {\bibfnamefont {W.}~\bibnamefont {Witczak-Krempa}},\
  }\href {\doibase 10.1007/jhep09(2015)091} {\bibfield  {journal} {\bibinfo
  {journal} {J. High Energ. Phys.}\ }\textbf {\bibinfo {volume} {2015}}
  (\bibinfo {year} {2015}),\ 10.1007/jhep09(2015)091},\ \Eprint
  {http://arxiv.org/abs/arXiv:1507.06997} {arXiv:1507.06997} \BibitemShut
  {NoStop}%
\bibitem [{\citenamefont {{Wang}}\ \emph {et~al.}(2021)\citenamefont {{Wang}},
  \citenamefont {{Ma}}, \citenamefont {{Cheng}},\ and\ \citenamefont
  {{Meng}}}]{YCWang2021DQCdisorder}%
  \BibitemOpen
  \bibfield  {author} {\bibinfo {author} {\bibfnamefont {Y.-C.}\ \bibnamefont
  {{Wang}}}, \bibinfo {author} {\bibfnamefont {N.}~\bibnamefont {{Ma}}},
  \bibinfo {author} {\bibfnamefont {M.}~\bibnamefont {{Cheng}}}, \ and\
  \bibinfo {author} {\bibfnamefont {Z.~Y.}\ \bibnamefont {{Meng}}},\
  }\href@noop {} {\bibfield  {journal} {\bibinfo  {journal} {arXiv e-prints}\
  ,\ \bibinfo {eid} {arXiv:2106.01380}} (\bibinfo {year} {2021})},\ \Eprint
  {http://arxiv.org/abs/2106.01380} {arXiv:2106.01380 [cond-mat.str-el]}
  \BibitemShut {NoStop}%
\bibitem [{\citenamefont {Nahum}(2020)}]{Nahum2020}%
  \BibitemOpen
  \bibfield  {author} {\bibinfo {author} {\bibfnamefont {A.}~\bibnamefont
  {Nahum}},\ }\href {\doibase 10.1103/PhysRevB.102.201116} {\bibfield
  {journal} {\bibinfo  {journal} {Phys. Rev. B}\ }\textbf {\bibinfo {volume}
  {102}},\ \bibinfo {pages} {201116} (\bibinfo {year} {2020})}\BibitemShut
  {NoStop}%
\bibitem [{\citenamefont {Ma}\ and\ \citenamefont {Wang}(2020)}]{RCMa2020}%
  \BibitemOpen
  \bibfield  {author} {\bibinfo {author} {\bibfnamefont {R.}~\bibnamefont
  {Ma}}\ and\ \bibinfo {author} {\bibfnamefont {C.}~\bibnamefont {Wang}},\
  }\href {\doibase 10.1103/PhysRevB.102.020407} {\bibfield  {journal} {\bibinfo
   {journal} {Phys. Rev. B}\ }\textbf {\bibinfo {volume} {102}},\ \bibinfo
  {pages} {020407} (\bibinfo {year} {2020})}\BibitemShut {NoStop}%
\bibitem [{\citenamefont {Zhao}\ \emph {et~al.}(2020)\citenamefont {Zhao},
  \citenamefont {Takahashi},\ and\ \citenamefont {Sandvik}}]{BWZhao2020PRL}%
  \BibitemOpen
  \bibfield  {author} {\bibinfo {author} {\bibfnamefont {B.}~\bibnamefont
  {Zhao}}, \bibinfo {author} {\bibfnamefont {J.}~\bibnamefont {Takahashi}}, \
  and\ \bibinfo {author} {\bibfnamefont {A.~W.}\ \bibnamefont {Sandvik}},\
  }\href {\doibase 10.1103/PhysRevLett.125.257204} {\bibfield  {journal}
  {\bibinfo  {journal} {Phys. Rev. Lett.}\ }\textbf {\bibinfo {volume} {125}},\
  \bibinfo {pages} {257204} (\bibinfo {year} {2020})}\BibitemShut {NoStop}%
\bibitem [{\citenamefont {Kuklov}\ \emph {et~al.}(2008)\citenamefont {Kuklov},
  \citenamefont {Matsumoto}, \citenamefont {Prokof'ev}, \citenamefont
  {Svistunov},\ and\ \citenamefont {Troyer}}]{Kuklov2008}%
  \BibitemOpen
  \bibfield  {author} {\bibinfo {author} {\bibfnamefont {A.~B.}\ \bibnamefont
  {Kuklov}}, \bibinfo {author} {\bibfnamefont {M.}~\bibnamefont {Matsumoto}},
  \bibinfo {author} {\bibfnamefont {N.~V.}\ \bibnamefont {Prokof'ev}}, \bibinfo
  {author} {\bibfnamefont {B.~V.}\ \bibnamefont {Svistunov}}, \ and\ \bibinfo
  {author} {\bibfnamefont {M.}~\bibnamefont {Troyer}},\ }\href {\doibase
  10.1103/PhysRevLett.101.050405} {\bibfield  {journal} {\bibinfo  {journal}
  {Phys. Rev. Lett.}\ }\textbf {\bibinfo {volume} {101}},\ \bibinfo {pages}
  {050405} (\bibinfo {year} {2008})}\BibitemShut {NoStop}%
\bibitem [{\citenamefont {Chen}\ \emph {et~al.}(2013)\citenamefont {Chen},
  \citenamefont {Huang}, \citenamefont {Deng}, \citenamefont {Kuklov},
  \citenamefont {Prokof'ev},\ and\ \citenamefont {Svistunov}}]{KChen2013}%
  \BibitemOpen
  \bibfield  {author} {\bibinfo {author} {\bibfnamefont {K.}~\bibnamefont
  {Chen}}, \bibinfo {author} {\bibfnamefont {Y.}~\bibnamefont {Huang}},
  \bibinfo {author} {\bibfnamefont {Y.}~\bibnamefont {Deng}}, \bibinfo {author}
  {\bibfnamefont {A.~B.}\ \bibnamefont {Kuklov}}, \bibinfo {author}
  {\bibfnamefont {N.~V.}\ \bibnamefont {Prokof'ev}}, \ and\ \bibinfo {author}
  {\bibfnamefont {B.~V.}\ \bibnamefont {Svistunov}},\ }\href {\doibase
  10.1103/PhysRevLett.110.185701} {\bibfield  {journal} {\bibinfo  {journal}
  {Phys. Rev. Lett.}\ }\textbf {\bibinfo {volume} {110}},\ \bibinfo {pages}
  {185701} (\bibinfo {year} {2013})}\BibitemShut {NoStop}%
\bibitem [{\citenamefont {{D'Emidio}}\ \emph {et~al.}(2021)\citenamefont
  {{D'Emidio}}, \citenamefont {{Eberharter}},\ and\ \citenamefont
  {{L{\"a}uchli}}}]{DEmidio2021}%
  \BibitemOpen
  \bibfield  {author} {\bibinfo {author} {\bibfnamefont {J.}~\bibnamefont
  {{D'Emidio}}}, \bibinfo {author} {\bibfnamefont {A.~A.}\ \bibnamefont
  {{Eberharter}}}, \ and\ \bibinfo {author} {\bibfnamefont {A.~M.}\
  \bibnamefont {{L{\"a}uchli}}},\ }\href@noop {} {\bibfield  {journal}
  {\bibinfo  {journal} {arXiv e-prints}\ ,\ \bibinfo {eid} {arXiv:2106.15462}}
  (\bibinfo {year} {2021})},\ \Eprint {http://arxiv.org/abs/2106.15462}
  {arXiv:2106.15462 [cond-mat.str-el]} \BibitemShut {NoStop}%
\bibitem [{\citenamefont {Shao}\ \emph {et~al.}(2016)\citenamefont {Shao},
  \citenamefont {Guo},\ and\ \citenamefont {Sandvik}}]{Shao2016}%
  \BibitemOpen
  \bibfield  {author} {\bibinfo {author} {\bibfnamefont {H.}~\bibnamefont
  {Shao}}, \bibinfo {author} {\bibfnamefont {W.}~\bibnamefont {Guo}}, \ and\
  \bibinfo {author} {\bibfnamefont {A.~W.}\ \bibnamefont {Sandvik}},\ }\href
  {\doibase 10.1126/science.aad5007} {\bibfield  {journal} {\bibinfo  {journal}
  {Science}\ }\textbf {\bibinfo {volume} {352}},\ \bibinfo {pages} {213}
  (\bibinfo {year} {2016})}\BibitemShut {NoStop}%
\bibitem [{sup()}]{suppl}%
  \BibitemOpen
  \href@noop {} {\bibinfo  {journal} {The implementation protocol of the
  nonequilibrium increment of the R\'enyi entanglement entropy and details of
  the numerical results with the finite size scaling with different boundary
  conditions of the entangling region}\ }\BibitemShut {NoStop}%
\bibitem [{\citenamefont {Sandvik}(1999)}]{Sandvik1999}%
  \BibitemOpen
\bibfield  {journal} {  }\bibfield  {author} {\bibinfo {author} {\bibfnamefont
  {A.~W.}\ \bibnamefont {Sandvik}},\ }\href {\doibase
  10.1103/PhysRevB.59.R14157} {\bibfield  {journal} {\bibinfo  {journal} {Phys.
  Rev. B}\ }\textbf {\bibinfo {volume} {59}},\ \bibinfo {pages} {R14157}
  (\bibinfo {year} {1999})}\BibitemShut {NoStop}%
\bibitem [{\citenamefont {Sylju\aa{}sen}\ and\ \citenamefont
  {Sandvik}(2002)}]{Syljuaasen2002}%
  \BibitemOpen
  \bibfield  {author} {\bibinfo {author} {\bibfnamefont {O.~F.}\ \bibnamefont
  {Sylju\aa{}sen}}\ and\ \bibinfo {author} {\bibfnamefont {A.~W.}\ \bibnamefont
  {Sandvik}},\ }\href {\doibase 10.1103/PhysRevE.66.046701} {\bibfield
  {journal} {\bibinfo  {journal} {Phys. Rev. E}\ }\textbf {\bibinfo {volume}
  {66}},\ \bibinfo {pages} {046701} (\bibinfo {year} {2002})}\BibitemShut
  {NoStop}%
\bibitem [{\citenamefont {Jarzynski}(1997)}]{Jarzynski1997}%
  \BibitemOpen
  \bibfield  {author} {\bibinfo {author} {\bibfnamefont {C.}~\bibnamefont
  {Jarzynski}},\ }\href {\doibase 10.1103/PhysRevLett.78.2690} {\bibfield
  {journal} {\bibinfo  {journal} {Phys. Rev. Lett.}\ }\textbf {\bibinfo
  {volume} {78}},\ \bibinfo {pages} {2690} (\bibinfo {year}
  {1997})}\BibitemShut {NoStop}%
\bibitem [{\citenamefont {Assaad}\ and\ \citenamefont
  {Herbut}(2013)}]{Assaad2013}%
  \BibitemOpen
  \bibfield  {author} {\bibinfo {author} {\bibfnamefont {F.~F.}\ \bibnamefont
  {Assaad}}\ and\ \bibinfo {author} {\bibfnamefont {I.~F.}\ \bibnamefont
  {Herbut}},\ }\href {\doibase 10.1103/PhysRevX.3.031010} {\bibfield  {journal}
  {\bibinfo  {journal} {Phys. Rev. X}\ }\textbf {\bibinfo {volume} {3}},\
  \bibinfo {pages} {031010} (\bibinfo {year} {2013})}\BibitemShut {NoStop}%
\bibitem [{\citenamefont {Nahum}\ \emph {et~al.}(2015)\citenamefont {Nahum},
  \citenamefont {Chalker}, \citenamefont {Serna}, \citenamefont {Ortu\~no},\
  and\ \citenamefont {Somoza}}]{Nahum2015}%
  \BibitemOpen
  \bibfield  {author} {\bibinfo {author} {\bibfnamefont {A.}~\bibnamefont
  {Nahum}}, \bibinfo {author} {\bibfnamefont {J.~T.}\ \bibnamefont {Chalker}},
  \bibinfo {author} {\bibfnamefont {P.}~\bibnamefont {Serna}}, \bibinfo
  {author} {\bibfnamefont {M.}~\bibnamefont {Ortu\~no}}, \ and\ \bibinfo
  {author} {\bibfnamefont {A.~M.}\ \bibnamefont {Somoza}},\ }\href {\doibase
  10.1103/PhysRevX.5.041048} {\bibfield  {journal} {\bibinfo  {journal} {Phys.
  Rev. X}\ }\textbf {\bibinfo {volume} {5}},\ \bibinfo {pages} {041048}
  (\bibinfo {year} {2015})}\BibitemShut {NoStop}%
\bibitem [{\citenamefont {Wang}\ \emph {et~al.}(2021)\citenamefont {Wang},
  \citenamefont {Zaletel}, \citenamefont {Mong},\ and\ \citenamefont
  {Assaad}}]{ZJWang2021}%
  \BibitemOpen
  \bibfield  {author} {\bibinfo {author} {\bibfnamefont {Z.}~\bibnamefont
  {Wang}}, \bibinfo {author} {\bibfnamefont {M.~P.}\ \bibnamefont {Zaletel}},
  \bibinfo {author} {\bibfnamefont {R.~S.~K.}\ \bibnamefont {Mong}}, \ and\
  \bibinfo {author} {\bibfnamefont {F.~F.}\ \bibnamefont {Assaad}},\ }\href
  {\doibase 10.1103/PhysRevLett.126.045701} {\bibfield  {journal} {\bibinfo
  {journal} {Phys. Rev. Lett.}\ }\textbf {\bibinfo {volume} {126}},\ \bibinfo
  {pages} {045701} (\bibinfo {year} {2021})}\BibitemShut {NoStop}%
\bibitem [{\citenamefont {Ma}\ and\ \citenamefont {He}(2019)}]{MaHe2019}%
  \BibitemOpen
  \bibfield  {author} {\bibinfo {author} {\bibfnamefont {H.}~\bibnamefont
  {Ma}}\ and\ \bibinfo {author} {\bibfnamefont {Y.-C.}\ \bibnamefont {He}},\
  }\href {\doibase 10.1103/PhysRevB.99.195130} {\bibfield  {journal} {\bibinfo
  {journal} {Phys. Rev. B}\ }\textbf {\bibinfo {volume} {99}},\ \bibinfo
  {pages} {195130} (\bibinfo {year} {2019})}\BibitemShut {NoStop}%
\bibitem [{\citenamefont {Gorbenko}\ \emph
  {et~al.}(2018{\natexlab{a}})\citenamefont {Gorbenko}, \citenamefont
  {Rychkov},\ and\ \citenamefont {Zan}}]{Gorbenko2018I}%
  \BibitemOpen
  \bibfield  {author} {\bibinfo {author} {\bibfnamefont {V.}~\bibnamefont
  {Gorbenko}}, \bibinfo {author} {\bibfnamefont {S.}~\bibnamefont {Rychkov}}, \
  and\ \bibinfo {author} {\bibfnamefont {B.}~\bibnamefont {Zan}},\ }\href
  {\doibase 10.1007/JHEP10(2018)108} {\bibfield  {journal} {\bibinfo  {journal}
  {Journal of High Energy Physics}\ }\textbf {\bibinfo {volume} {2018}},\
  \bibinfo {pages} {108} (\bibinfo {year} {2018}{\natexlab{a}})}\BibitemShut
  {NoStop}%
\bibitem [{\citenamefont {Gorbenko}\ \emph
  {et~al.}(2018{\natexlab{b}})\citenamefont {Gorbenko}, \citenamefont
  {Rychkov},\ and\ \citenamefont {Zan}}]{Gorbenko2018II}%
  \BibitemOpen
  \bibfield  {author} {\bibinfo {author} {\bibfnamefont {V.}~\bibnamefont
  {Gorbenko}}, \bibinfo {author} {\bibfnamefont {S.}~\bibnamefont {Rychkov}}, \
  and\ \bibinfo {author} {\bibfnamefont {B.}~\bibnamefont {Zan}},\ }\href
  {\doibase 10.21468/SciPostPhys.5.5.050} {\bibfield  {journal} {\bibinfo
  {journal} {SciPost Phys.}\ }\textbf {\bibinfo {volume} {5}},\ \bibinfo
  {pages} {50} (\bibinfo {year} {2018}{\natexlab{b}})}\BibitemShut {NoStop}%
\end{thebibliography}%

\setcounter{figure}{0}
\setcounter{page}{1}
\renewcommand{\theequation}{S\arabic{equation}}
\renewcommand{\thefigure}{S\arabic{figure}}

\newpage
\newpage

\onecolumngrid

\section*{Supplemental Material}

\subsection{The protocol of nonequilibrium increment method}

The protocol of nonequilibrium increment measurement is listed below:
\begin{enumerate}
	\item We prepare the stochastic series expansion (SSE) QMC~\cite{Sandvik1999,Syljuaasen2002}  on one replica and storage the thermalized configuration.
	\item We send the thermalized configuration  to $K$ parallel processes and make two copies of the replicas. Then the two replicas will be thermalized again according to  $Z^{(2)}_{A}(\lambda(t_{i}))$. For process $k$ we set the initial value of $\lambda$ to be $\lambda(t_{i})=k\Delta$. The value of $\lambda$ determines the probability of the sites in $A$ (the entangling region) joining or leaving the topology trace. 
	\item We then perform a MC sweep which consists of several MC steps. Each step is defined as follows.
	\begin{itemize}
		\item Each site in $A$ can choose whether to stay or leave the topology trace based on  probabilities regulated by $\lambda$. Here
		\begin{equation}
		P_{\text{join}}=\min\{\frac{\lambda}{1-\lambda},1\} \quad P_{\text{split}}=\min\{\frac{1-\lambda}{\lambda},1\}
		\end{equation}
		\item For each spin, after the decision is made, accept it if the boundary condition is still satisfied after change of the topology.
		\item After the trace structure is determined, carry out a diagonal update and loop update to flip the spins and operator in the SSE configuration.
	\end{itemize}
	\item Record the value of observable $\langle \frac{g_{A}(\lambda(t_{m+1}),N_{B}(t_{m}))}{g_{A}(\lambda(t_{m}),N_{B}(t_{m}))}\rangle$. Increase $t$ by $\Delta t$ until $t$ reaches the value of $(k+1)\Delta$ and go back to step 3.
	\item In the end, we collect the observable from all the processes and sum them to get the total entropy.
\end{enumerate}

\subsection{The choice of entangling region $A$}
\begin{figure*}[htbp]
	\centering
	\includegraphics[width=\columnwidth]{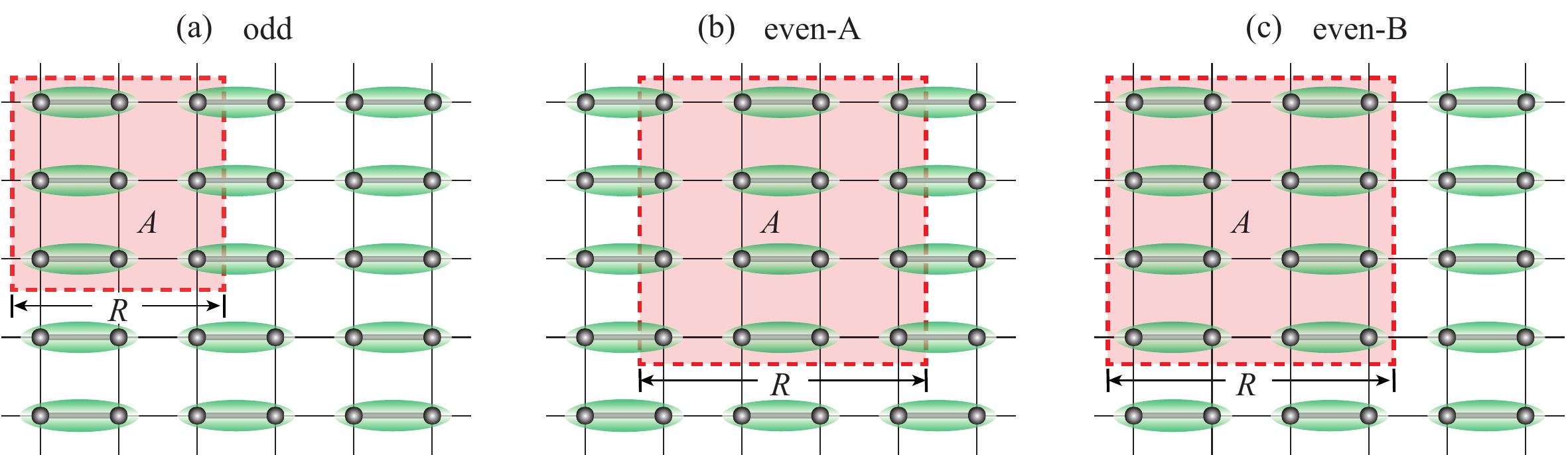}
	\caption{The types of entangling regions. (a) odd: the entangling region $A$ has an odd linear size, e.g. $R=3$.(b) even-A: the entangling region $A$ has an even linear size, e.g. $R=4$, but cuts the strong dimer bonds at two parallel sides.(c) even-B: the entangling region $A$ has an even linear size, e.g. $R=4$, without cutting any strong dimer bonds.}
	\label{fig:figs1}
\end{figure*}

In this section, we discuss how to choose the entangling region $A$ to give the accurate finite size scaling behavior of the EE. As discussed in the calculation of disorder operator in Ref. \cite{YCWang2021DQCdisorder}, there are three types of resions: odd, even-A, and even-B, which are shown in Fig.~~\ref{fig:figs1}.

\begin{figure}[htbp]
	\centering
	\includegraphics[width=0.55\columnwidth]{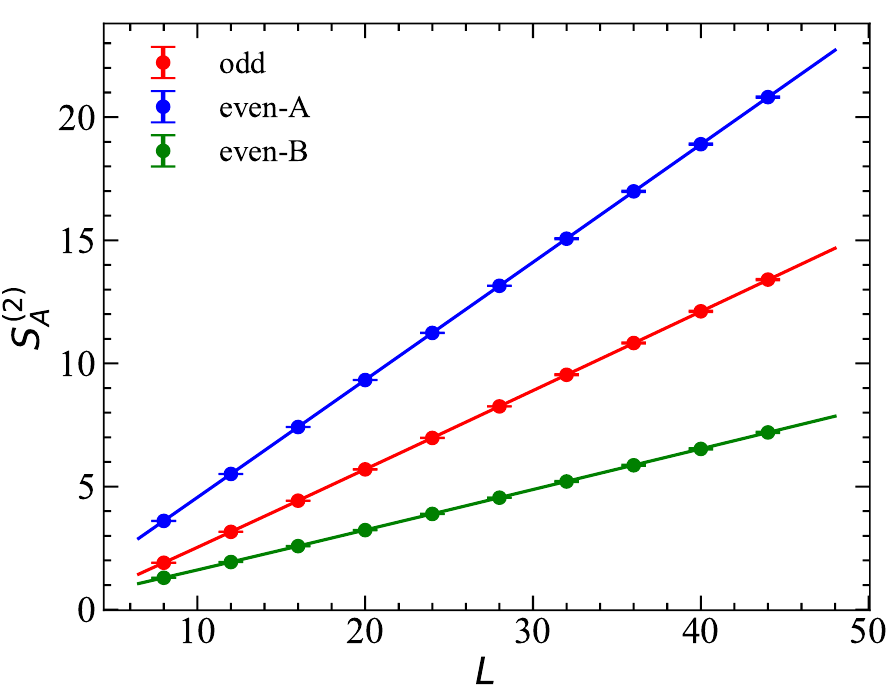}
	\caption{Second R\'enyi entropy for the three types of entangling regions of the $L\times L$ square lattice $H_{J_1-J_2}$ model. We use the relation $S_{A}^{(2)}=aL-s\ln(L)+b$ to perform the curve fitting. We obtain that
		the fitting result for region odd is  $S_{A}^{(2)}=0.324(1)L-0.103(5)\ln(L)-471(7)$,
		the fitting result for region even-A is  $S_{A}^{(2)}=0.479(1)L-0.034(4)\ln(L)-0.158(7)$,
	    and the fitting result for region even-B is  $S_{A}^{(2)}=0.168(1)L-0.081(4)\ln(L)-0.124(7)$.}
	\label{fig:figs2}
\end{figure}

The second R\'enyi EE of the $H_{J_1-J_2}$ model with three different choices of the entangling region is detected with the noneuqilibrium increment method and the results are shown on Fig.~\ref{fig:figs2}. We find that region even-B gives $s=0.081(4)$ which is in agreement with the results obtained in O(3) models~\cite{KallinJS2014,Helmes2014,JRZhao2020}, representing the central charge of the O(n) CFT. While for region odd and even-A, the obtained values of $s$ both deviate from the O(3) results. The fitting results indicate that cutting no strong dimer bonds(even-B) will bests hold the universality of the logarithmic correction coefficient $s$, while cutting dimer bonds will cause a shift of the coefficient.

\begin{figure}[ht]
\centering
\includegraphics[width=0.55\columnwidth]{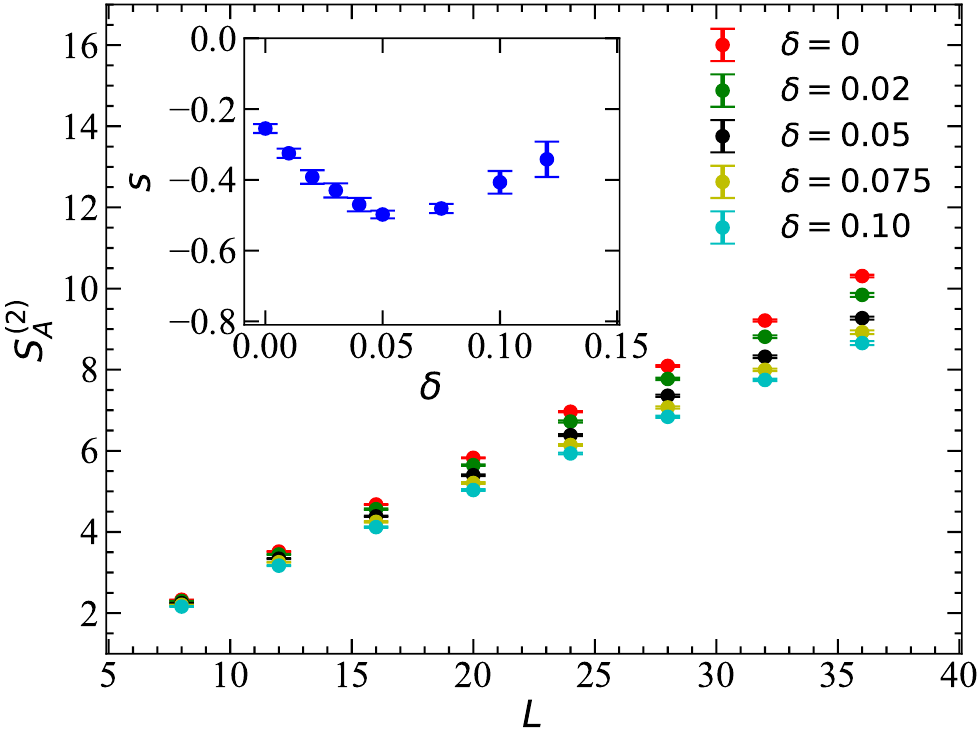}
\caption{The second R\'enyi entanglement entropy $S_{A}^{(2)}$ versus the system size $L$ for the $H_{J-Q_3}$ model with the entangling region A chosen to be the type of even-B for different pinning fields. The inset shows the magnitude of pinning field $\delta$ versus the obtained coefficient $s$. The converged value of $s$ at $\delta\sim 0.05$ is observed. For all the $\delta$ values, we find $s<0$.}
\label{fig:figs3}
\end{figure}

For the $H_{J-Q_3}$ model at the DQC, the strong dimer bonds are not naturally arranged as shown in Fig.~\ref{fig:figs1}. In fact the bonds can be generated from both $x$ and $y$ directions. So in this case we are not able to distinguish between the three types of entangling regions because we do not know how many dimer bonds the boundary cuts. However by introducing a small pinning field $J_2=J+\delta/L$ in the columnar dimer arrangement of $H_{J_1-J_2}$ added to $J$ bonds, the dimer bonds are expected to arrange in the direction of $J$ because bonds prefer lower energy. 

With a pinning field, the second R\'enyi entanglement entropy $S_{A}^{(2)}$ with $A$ chosen to be the type of even-B can be measured and the results are shown on Fig.~\ref{fig:figs3}. We find that the small pinning field will also cause a shift of the value of $s$. However when $\delta$ is around 0.05, the obtained $s$ seems to become flat and then gradually decreases as $\delta$ is further magnified, therefore $\delta=0.05$ senses the correct length scale that on one hand generated the prefered VBS order but on the other does not affect the finite extrapolation of the intrinsic DQC scaling towards the thermodynamic limit, with the system sizes we studied here. In the main text, we present the DQC $S^{(2)}_A(L)$ analysis with $\delta=0.05$. We have also notice that with all the choices of $\delta$ (including $\delta=0$), the coefficient logarithmic correction correction $s$ is always negative, {which indicates the robustness of our our findings under small perturbations.}

\vspace{64pt}

\end{document}